\documentclass[amsmath,superscriptaddress,aps,pra,twocolumn]{revtex4-2}

\usepackage{graphicx}
\usepackage{epsfig}        
\usepackage{dcolumn}
\usepackage{bm}

\usepackage{amsmath}
\usepackage{amssymb}
\usepackage{mathtools}

\usepackage{array}
\usepackage{multirow}
\usepackage{booktabs}
\usepackage{tabularx}
\newcommand{\br}{}
\newcommand{\mr}{}

\usepackage{chemformula}
\usepackage{siunitx}
\usepackage{gensymb}

\usepackage{braket}

\usepackage{url}
\usepackage{xcolor}
\usepackage[caption=false]{subfig}

\graphicspath{{images/}}
\usepackage[colorlinks=true,allcolors=blue]{hyperref}

\begin{document}
\title{Circular V-grooves on single-crystal gold: optical properties and sensing feasibility}

\author{Amos S. Kiyumbi}
\email{amos.kiyumbi@udsm.ac.tz}
\affiliation{Department of Physics, Mathematics and Informatics, Dar es Salaam University College of Education, P.O.
Box 2329, Dar es Salaam, Tanzania}
\affiliation{Ulm University, Institute of Experimental Physics, Albert-Einstein-Allee 11, 89069 Ulm, Germany}
\affiliation{Department of Physics, Stellenbosch University, Private Bag X1, Matieland 7602, South Africa}

\begin{abstract}
Single-crystal Au(111) microplates provide an ultra-smooth, low-defect platform for reproducible plasmonic nanocavities. Here we realize reflection-mode whispering-gallery metasurfaces comprising periodic arrays of circular V-groove cavities milled into optically thick Au microplates and characterize their visible--near-IR response. The measured spectra exhibit narrow, depth-tunable Fano-like reflectance minima with weak azimuthal dependence, reproduced by quarter-cell finite-element modeling consistent with strong gap-surface-plasmon confinement. For refractometric sensing, simulations yield bulk sensitivities up to $\sim 598$~nm~RIU$^{-1}$ and figures of merit up to $\sim 33$ (hexagonal lattice). As a model-based illustration, a functionalized architecture targeting \textit{Plasmodium falciparum}
lactate dehydrogenase (PfLDH), a common malaria target of rapid diagnostic tests, gives an estimated limit of detection of $0.016$~nM ($\sim 0.56$~ng~mL$^{-1}$) under the adopted noise floor.
\end{abstract}
\maketitle

\section{Introduction}
Metasurfaces provide a versatile route to engineer spectral responses within ultrathin, subwavelength platforms, enabling compact implementations of nonlinear optics and sensing~\cite{kildishev2013planar,yu2014flat}. Within this landscape, plasmonic metasurfaces are attractive for refractometric biosensing because they concentrate electromagnetic energy near metal interfaces, yielding strong spectral signatures that are sensitive to the local dielectric environment~\cite{homola2008surface,anker2008biosensing}. A particularly effective geometry is the metal--insulator--metal (MIM) groove, which supports gap-surface-plasmon ($gsp$) modes with large effective index and strong field confinement near the groove bottom~\cite{bozhevolnyi2006effective,smith2015gap}. In practice, however, achieving narrow and reproducible resonances in metallic nanostructures remains challenging because ohmic loss and fabrication-induced roughness broaden spectral features and can reduce device-to-device consistency~\cite{smith2015gap,duan2022recent}. Single-crystal gold offers a compelling route to mitigate extrinsic scattering and damping, providing atomically smooth surfaces and reduced grain-boundary losses compared with conventional polycrystalline films~\cite{kaltenecker2020mono,radha2012giant}.\\

Plasmon guiding and confinement in V-grooves has been extensively studied in the context of channel plasmon polaritons and related wedge/channel modes, where subwavelength grooves can laterally compress surface plasmon fields below the diffraction limit while retaining propagating behavior~\cite{bozhevolnyi2005channel,moreno2006channel}. For sufficiently narrow gaps, the strongly increased modal effective index and the localization of energy near the groove bottom lead to enhanced absorption and scattering signatures that are highly sensitive to nanoscale geometry (width, depth, sidewall angle) and to the local dielectric environment~\cite{bozhevolnyi2006channel,dintinger2009channel}. These characteristics motivate V-groove architectures as building blocks for compact resonators and metasurfaces, particularly when the metal quality (surface roughness, crystallinity) is high enough to prevent extrinsic broadening from dominating the resonance linewidth.\\

Curving a linear V-groove into a circular nanocavity creates a compact resonator in which circulating $gsp$ form whispering-gallery-like standing-wave patterns along the circumference~\cite{vesseur2009modal}. Under normal illumination, arrays of such circular V-groove resonators, similar to other plasmonic resonators, have been explored for many applications~\cite{meinzer2014plasmonic,kasani2019review}. However, under oblique illumination, $gsp$ modes in the circular V-groove can interfere between multipolar contributions and access radiatively suppressed regimes, commonly discussed in the broader context of optical anapoles~\cite{savinov2019optical,miroshnichenko2015nonradiating,yezekyan2022anapole}. Most prior demonstrations, however, employ thin-film or polycrystalline metals, where microstructural disorder can obscure intrinsic cavity physics and complicate quantitative comparison between experiment and modeling~\cite{duan2022recent,kaltenecker2020mono}.

In addition to spectral selectivity, linear V-groove and groove-ring resonators are increasingly relevant for sensing platforms because they can provide sharp resonances and steep spectral slopes while maintaining a compact footprint, which is advantageous for miniaturized readout and reduced sample volume. In practical biosensing, performance is typically communicated via sensitivity and linewidth-normalized metrics, together with limit-of-detection estimates that depend on both the intrinsic optical response and the readout strategy (wavelength versus intensity interrogation, noise floor, drift, and resonance-extraction uncertainty)~\cite{piliarik2009surface,mayer2011localized}.\\

Beyond establishing the optical response, we numerically explore regimes relevant to radiatively suppressed resonances under oblique incidence. In this context, our simulations indicate an anapole-like excitation characterized by a pronounced reflection minimum accompanied by enhanced stored electromagnetic energy within the cavity~\cite{yezekyan2022anapole,savinov2019optical}; in the present work, we report an anapole-like response based on qualitative near- and far-field signatures extracted from the simulations, and a full multipolar decomposition is left for future study, where experiments will be compared with simulations. Finally, building on the bulk spectral response, we assess the feasibility of surface refractometric sensing and model the detection of \textit{Plasmodium falciparum} (Pf), a malaria biomarker spiked in a buffer solution, using intensity interrogation.

The global malaria burden remains high due to drug-resistant parasites, insecticide-resistant mosquitoes, funding shortages, climate-driven environmental changes, and weakened health systems, with the greatest impact in sub-Saharan Africa~\cite{WHO2025WMR, platon2026world}. This drives ongoing development of sensitive, robust diagnostics. PfLDH, a glycolytic enzyme in the \textit{Plasmodium falciparum} malaria parasite and a common rapid diagnostic test target conserved across human malaria species, shows performance and practical LOD values that strongly depend on test format, target epitope, and sample conditions~\cite{yerlikaya2022dual}. In this context, plasmonic biosensors (here theoretically modeled) are attractive because they can, in principle, provide label-free quantitative readout at low analyte levels, but the clinically relevant benchmark is ultimately determined by biomarker biology (release kinetics, clearance, species specificity) and the measurement matrix (e.g., whole-blood lysate).\\

In this work, we introduce a single-crystal plasmonic cavity metasurface platform that combines experimental spectral selectivity with validated electromagnetic modeling and sensing-relevant figures of merit. We fabricate periodic arrays of circular V-groove whispering-gallery nanocavities in optically thick Au(111) microplates and measure depth-tunable, Fano-like reflection minima that exhibit weak azimuthal dependence. Quarter-cell finite-element simulations reproduce the principal spectral trends and reveal strong gap-surface-plasmon confinement at the groove bottom and edges. Building on this validated response, we quantify refractometric performance for square and hexagonal lattices (bulk sensitivity and figure of merit) and provide a model-based biosensing case study for PfLDH to establish feasibility and key scaling constraints.

\section{Theory}
Linear V-shaped grooves, a simple class of MIM waveguides etched into a plasmonic metal surface, support guided plasmonic modes known as gap surface plasmons~\cite{bozhevolnyi2006effective, bozhevolnyi2008scaling} as stated in the previous section. These modes arise from strongly coupled surface plasmon polaritons confined within the dielectric-filled groove. A common characteristic of $gsp$ modes is their longitudinal nature and spatial localization at the metal edges or at the groove bottom; they are therefore also referred to as channel plasmons~\cite{novikov2002channel, bozhevolnyi2005channel}. These modes can propagate either along the groove axis or normal to the groove sidewalls.

The general dispersion relation for the waveguide modes, including the $gsp$ modes is given by~\cite{economou1969surface, bozhevolnyi2008scaling}
\begin{align}
\tanh\left(k_d \frac{w}{2}\right) = \frac{- \varepsilon_d k_m}{\varepsilon_m k_d}, \quad
k_{(m,d)} = \sqrt{k_{gsp}^2 - \varepsilon_{(m,d)} k_0^2},
\label{wave_guide}
\end{align}
where $w$ is the width of the dielectric gap between the two metal surfaces, $\varepsilon_{(m,d)}$ denote the permittivities of the metal ($m$) and the dielectric ($d$), $k_{gsp}$ is the propagation constant of the $gsp$ mode, and $k_0 = 2\pi/\lambda$ is the free-space wavenumber of the incident electromagnetic wave. For a moderately wide groove with $w > (\lambda \varepsilon_d)/(\pi |\varepsilon_m|)$ such that $\left|\frac{-2\varepsilon_d}{w \varepsilon_m}\right| < k_0$, the dispersion relation of $gsp$ modes supported by V-grooves, $k_{gsp}$, can be approximated as~\cite{bozhevolnyi2008scaling}
\begin{equation}
k_{gsp} \approx k_0 \sqrt{\varepsilon_d + \frac{2\varepsilon_d \sqrt{\varepsilon_d - \varepsilon_m}}{k_0 w (-\varepsilon_m)}}.
\label{kgps_linear}
\end{equation}
From Eq.~\eqref{kgps_linear} it follows that $gsp$ modes in linear grooves exist at all wavelengths and that the effective refractive index, $n_{eff} = k_{gsp}/k_0$, increases as the groove width decreases toward the groove bottom. The increase in $n_{eff}$ near the bottom of the groove is associated with enhanced ohmic losses (i.e., reduced propagation length and stronger evanescent decay into the dielectric, $\delta_{pen}:= (\lambda/2\pi)(n^2_{eff} - 1)^{-0.5}$~\cite{bozhevolnyi2007channelling}), which in turn leads to strong electromagnetic field confinement and enhancement at the groove bottom~\cite{bozhevolnyi2006effective, bozhevolnyi2008scaling, smith2015gap}.

By bending the linear V-groove into a circular geometry, a circular V-groove resonator is formed, which constitutes the meta-atom of the $wg$-metasurface considered in this work. The optical properties of a metasurface composed of such circular V-grooves are therefore intimately linked to the excitation of $gsp$ modes and can be further interpreted in terms of more complex resonances known as whispering-gallery modes ($wg$ modes)~\cite{vesseur2009modal, vesseur2010broadband}. These $wg$ modes correspond to circulating $gsp$ waves that form standing-wave patterns along the circumference of the V-shaped cavity. Similar to other plasmonic resonators, the optical response of $wg$ resonators can be tuned via the groove geometry (cavity diameter, width, and depth), the surrounding dielectric environment, and the polar angle of incidence of the illumination~\cite{gonccalves2020plasmonic}. A periodic array of circular V-cavities in a 100~nm-thick silver film has been reported to support anapole modes when the angle of incidence of the illumination exceeds $20\degree$~\cite{li2014excitation}. Furthermore, circular V-groove resonators (either isolated or arranged in arrays) have been proposed as candidates for frequency-selective surfaces~\cite{zhang2011continuous}, as versatile optical nanoantennas~\cite{vesseur2011plasmonic}, and as localized surface plasmon resonance sensors~\cite{Vesseur2011PhD}. In his Ph.D.\ thesis, Ernst Jan R.\ Vesseur~\cite{Vesseur2011PhD} additionally demonstrated an ultrasmall plasmonic whispering-gallery laser, exhibiting far-field emission with a distinct angular dependence governed by the azimuthal mode structure of the cavity.

\section{Materials and Methods}
\subsection{Metasurface fabrication}
Gold(III) chloride trihydrate and tetraoctylammonium bromide (ToABr) were employed as precursors for the synthesis of single-crystalline Au(111) microplates~\cite{radha2012giant}. The former was obtained from Sigma-Aldrich, the latter from Alfa Aesar, and both reagents were used as received, without further purification. The Au(111) synthesis method is based on a straightforward air-thermolysis process that involves the phase transfer of \ch{(AuCl4)-} ions from an aqueous phase to an organic toluene phase mediated by ToABr. Specifically, 0.059~g of Au(III) chloride trihydrate was dissolved in 5~mL of Milli-Q water to yield a 25~mM aqueous solution. A 50~mM ToABr solution in toluene was prepared by dissolving 0.137~g of ToABr in 5~mL of toluene. The two precursor solutions were then combined in a 1:2 volume ratio (for example, 450~$\mu$L of the Au(III) chloride trihydrate solution with 900~$\mu$L of the ToABr solution) and stirred to promote phase transfer. After completion of the phase transfer, the resulting orangish-red organic phase was drop-cast onto a cleaned glass substrate. The substrate was subsequently placed in an oven at 130~$\degree$C for 5~days. Well-defined gold microplates with hexagonal and triangular morphologies were observed after 2--3~days of thermal treatment.

\begin{figure*}[t!]
\begin{minipage}{0.49\linewidth}
\subfloat[]{\includegraphics[width=1\textwidth]{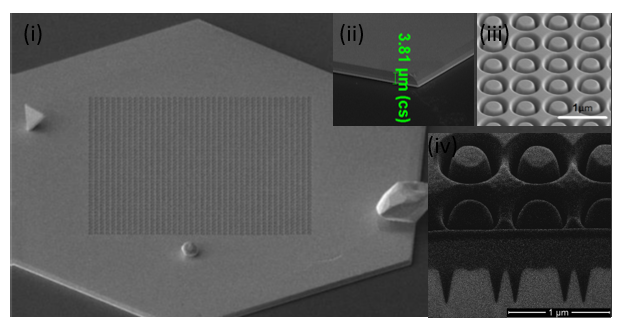}}
\end{minipage}%
\begin{minipage}{0.49\linewidth}
\subfloat[]{\includegraphics[width=1\textwidth]{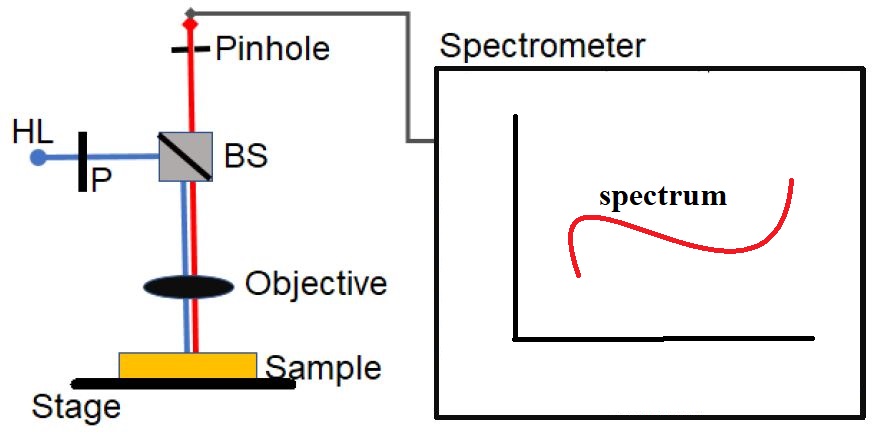}}
\end{minipage}\par\medskip
\caption{\label{SEM-image} The fabricated metasurface and its optical probing setup. (a)(i) The $wg$-metasurface ($P = 700$~nm, $D = 500$~nm, and $z = 300$~nm) on a hexagonally shaped Au(111) template. (a)(ii) SEM image of the Au(111) microplate ($3.81~\mu$m thick). (a)(iii) Close-up SEM image of the fabricated metasurface. (a)(iv) Cross-sectional view of the metasurface revealing the etch depth of the grooves. (b) Schematic drawing of the optical probing setup used. HL: halogen lamp; BS: beam splitter.}
\end{figure*}

The Au(111) microplates were then transferred and cleaved onto a cleaned silicon oxide substrate, after which arrays of resonators ($40\times 40$) were defined by focused Ga$^+$ ion beam (FIB) milling using a Helios NanoLab 600 (FEI), operated at 28~pA and 30~kV under normal incidence, to fabricate the desired metasurface. The metasurfaces were illuminated using an external broadband halogen white-light source (Ocean Optics HL-2000-FHSA). Reflection spectra were acquired with an exposure time of 10~ms and averaged over ten consecutive measurements. Spectral analysis of the reflected light was carried out using an Avantes AvaSpec SensLine spectrometer coupled to a WITec SNOM microscope and equipped with a Zeiss Achrostigmat objective (20$\times$, NA = 0.4). A representative SEM image of a $wg$-metasurface is presented in Fig.~\ref{SEM-image}(a), and a schematic of the optical interrogation scheme is shown in Fig.~\ref{SEM-image}(b). The resonance wavelength and linewidth were extracted from each reflection spectrum by identifying the minimum of the normalized reflectance and determining the corresponding full width at half maximum (FWHM) around this dip, with the resonance position optionally refined via spline fitting to reduce discretization noise.

\subsection{Finite-element method (FEM)}
\begin{figure*}[t!]
\centering
\includegraphics[width=\linewidth]{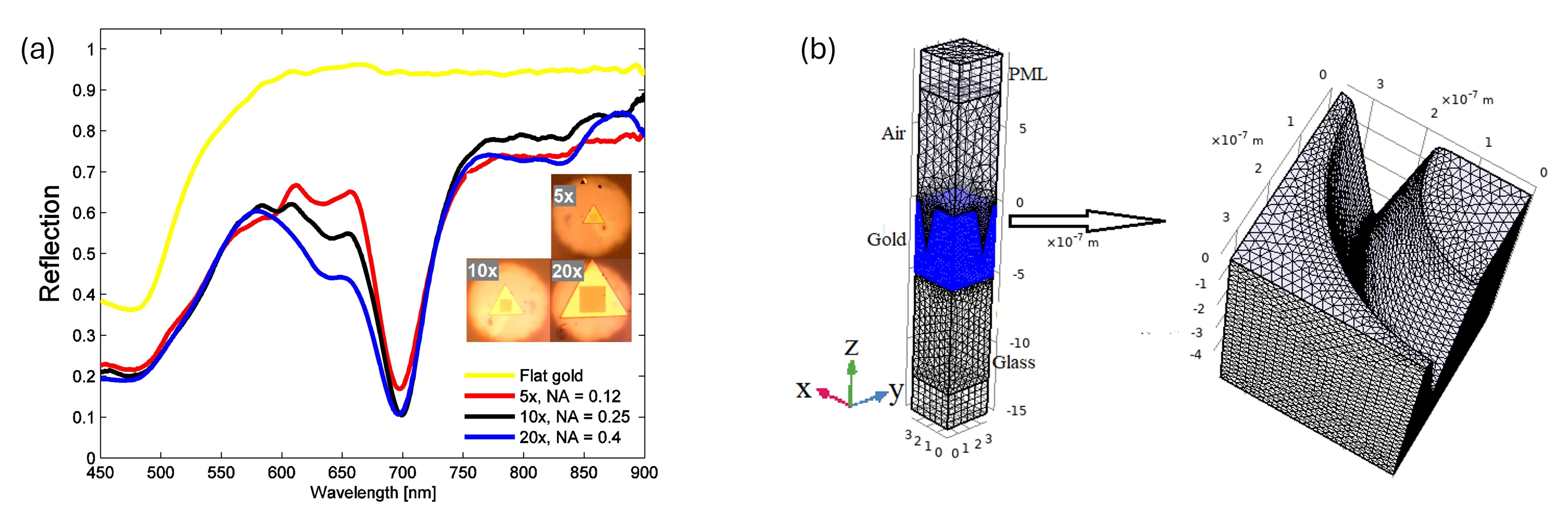}
\caption{\label{fig:appendix_spot_and_mesh}
Experimental interrogation area and FEM model geometry/meshing.
(a) Normalized reflectance spectra acquired from the metasurface using K\"ohler illumination for three microscope objectives (5$\times$, NA = 0.12; 10$\times$, NA = 0.25; 20$\times$, NA = 0.40), together with a reference spectrum from unstructured (flat) Au. The inset micrographs show the corresponding widefield illumination spots on the Au microplate, demonstrating that increasing magnification reduces the interrogated area and improves spatial selectivity for the patterned region. The finite numerical aperture also introduces an angular spread of incident and collected wave vectors, which can modestly broaden and reshape narrow resonances relative to idealized plane-wave excitation.
(b) Finite-element model used to compute the reflection response of the periodic metasurface. The left panel shows a quarter-unit-cell computational domain comprising air, Au, and glass, terminated by perfectly matched layers (PMLs) along the propagation direction to suppress boundary reflections; the geometry is discretized using a refined tetrahedral mesh in and around the nanocavity. The right panel shows a magnified view of the meshed V-groove profile, highlighting the mesh refinement required to accurately resolve the subwavelength groove width, depth, and sidewalls that govern $gsp$ confinement.}
\end{figure*}

The $wg$-metasurface was designed in COMSOL Multiphysics, a finite-element solver used to analyze wave-optics problems in physics and engineering. A unit resonator was designed using Floquet periodic boundary conditions to calculate the reflection response of the periodic nanogrooves forming the metasurface. Parameters used to design a unit cell of the $wg$-metasurface are shown in Appendix Fig.~\ref{design_appendix}. These include a circumference diameter $D = 500$~nm, groove depth $z = 300$~nm, a cavity width $w = 140$~nm (before applying fillets) and 160~nm (after applying fillets), gold thickness $t_{Au}:= z + 150$~nm (i.e., $\sim 450$--500~nm), and minor radii (fillets) $r_1 = 20$~nm and $r_2 = 10$~nm describing the curvature of the bottom and top corners of the groove, respectively. The quarter-unit-cell computational domain used for normal-incidence plane-wave excitation is shown in Fig.~\ref{fig:appendix_spot_and_mesh}(b).

Although the fabricated Au microplates are optically thick ($\sim 3.8~\mu$m), the gold thickness in the unit-cell model was truncated to reduce computational cost. We verified that this approximation does not affect the reported spectral features by performing a thickness-convergence study: increasing the simulated Au thickness beyond $\sim 500$~nm produced negligible changes in the resonance wavelength, linewidth (FWHM), and reflection-minimum depth over the wavelength range of interest. This confirms that the simulated structure is effectively optically thick and that the resonant response is governed primarily by the V-groove cavity geometry and near-field confinement rather than by finite-substrate-thickness effects.

Two lattice structures, square and hexagonal, are studied, and their spectra are compared. In both cases, the model geometry is discretized into finite elements formed by fine 3D meshing, and the frequency-domain Maxwell equations are solved numerically. The Johnson and Christy experimental data were used to incorporate the frequency-dependent permittivity $\varepsilon_m$ of gold~\cite{johnson1972optical} into the model. The refractive indices for a glass substrate, air, and deionized water were taken as 1.52, 1.00, and 1.333, respectively. To avoid undesired numerical reflections back into the interior of the computational region, the top and bottom domains that define the light-propagation direction are terminated by strongly absorbing perfectly matched layers. Moreover, due to the rotational symmetry of the circular cavities and the symmetry of the incident field, using perfect electric conductor (PEC) and perfect magnetic conductor (PMC) symmetry boundary conditions, quarter-cell FEM calculations reproduce full-cell results for normal plane-wave illumination. Therefore, a quarter cell was used to reproduce the experimental spectra. The COMSOL model used in this regard is shown in Fig.~\ref{fig:appendix_spot_and_mesh}(b).


\subsection{Device geometry and illumination}
The whispering-gallery metasurface ($wg$-metasurface) comprises a two-dimensional periodic array of circular V-groove nanocavities milled into an optically thick single-crystal Au microplate. Each unit cell contains a single circular groove of outer diameter $D$ and groove opening width $w$, with a nominal groove depth $z$ measured from the top Au surface to the groove bottom. The cavities are arranged either on a square lattice with period $p$ or on a hexagonal lattice with lattice constant $p$, enabling a controlled comparison of lattice symmetry while keeping the cavity geometry fixed. In the fabricated samples, arrays of $40\times 40$ resonators (overall lateral size $\sim 50~\mu$m) were patterned to ensure that the illuminated region was dominated by the periodic structure. Unless otherwise stated, the reference design parameters are $D=500$~nm, $w=160$~nm, and $p=700$~nm, with the cavity depth varied to tune the resonance. All optical characterization and simulations were performed over the visible--near-infrared spectral range ($\sim 400$--900~nm), for which the Au thickness ($\sim 3.8~\mu$m) is effectively optically thick.
\begin{figure*}[t!]
\begin{minipage}{.5\linewidth}
\centering
\subfloat[]{\includegraphics[width=1.0\textwidth]{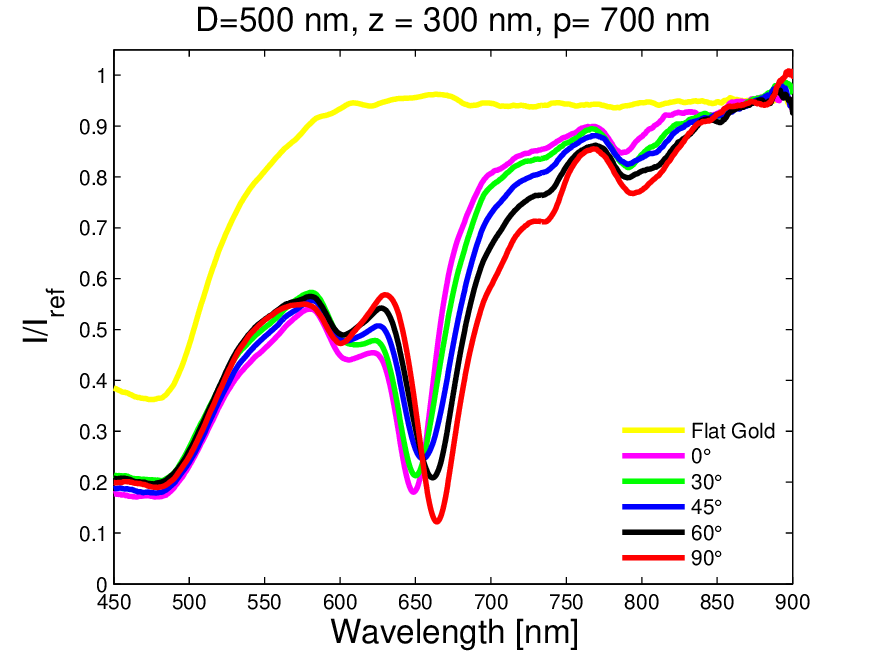}}
\end{minipage}%
\begin{minipage}{.5\linewidth}
\centering
\subfloat[]{\includegraphics[width=1.0\textwidth]{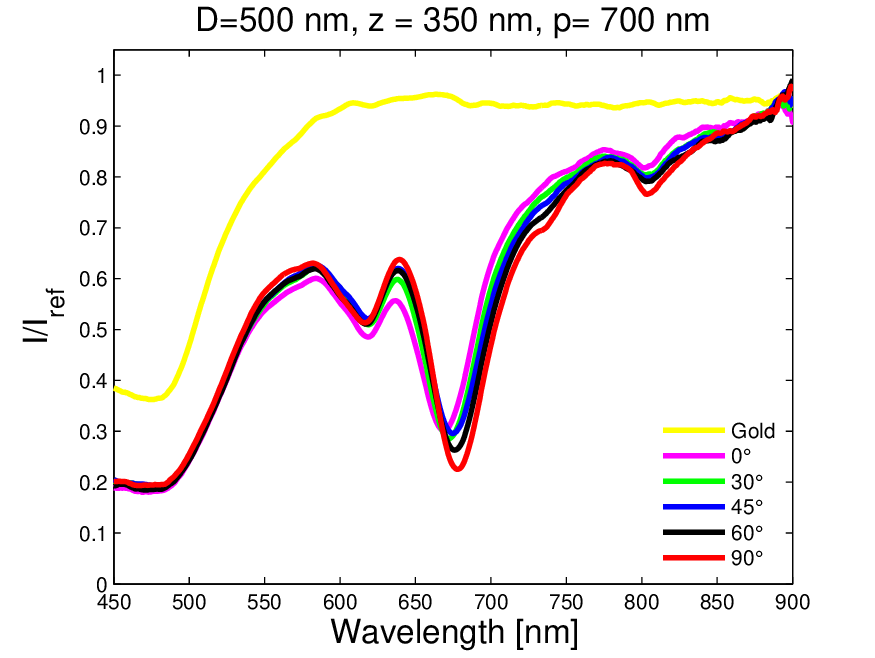}}
\end{minipage}\par\medskip
\begin{minipage}{.5\linewidth}
\centering
\subfloat[]{\includegraphics[width=1.0\textwidth]{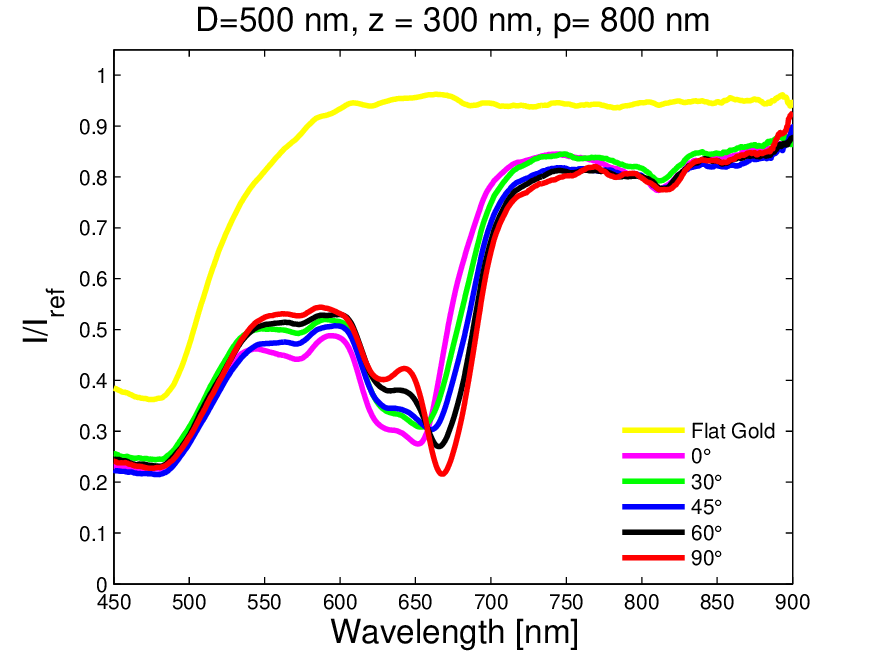}}
\end{minipage}%
\begin{minipage}{.5\linewidth}
\centering
\subfloat[]{\includegraphics[width=1.0\textwidth]{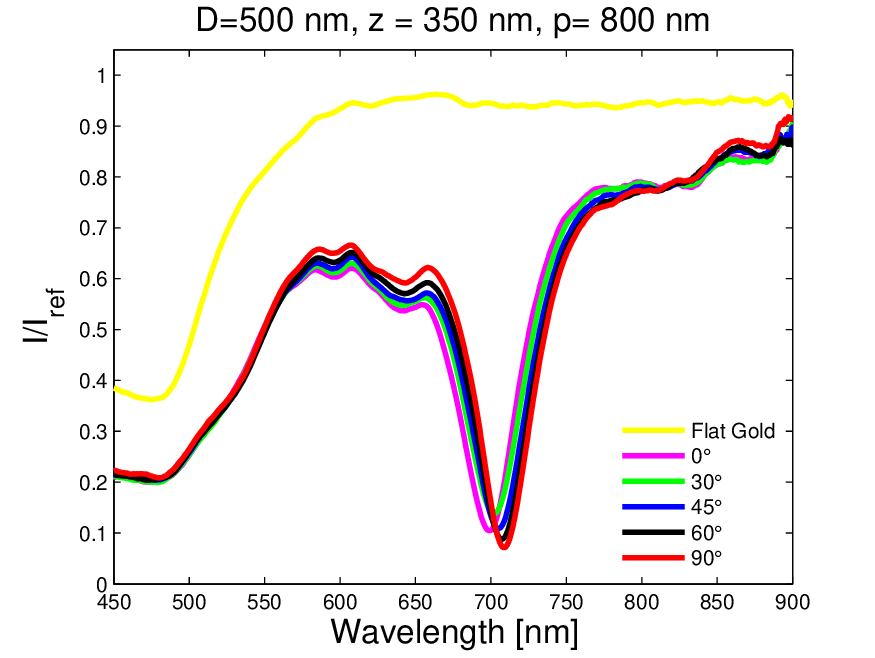}}
\end{minipage}\par\medskip
\caption{\label{appendix} Azimuthal-angle dependence of the measured reflection spectra for circular V-groove metasurfaces with fixed cavity diameter $D=\SI{500}{\nano\metre}$ and four $(p,z)$ combinations. Panels show: (a) $p=\SI{700}{\nano\metre}$, $z=\SI{300}{\nano\metre}$; (b) $p=\SI{700}{\nano\metre}$, $z=\SI{350}{\nano\metre}$; (c) $p=\SI{800}{\nano\metre}$, $z=\SI{300}{\nano\metre}$; and (d) $p=\SI{800}{\nano\metre}$, $z=\SI{350}{\nano\metre}$. Each subplot compares spectra acquired at different azimuthal viewing angles $\phi$ (as indicated in the legend) with the response of unstructured (flat) Au (yellow). Across all geometries, the principal Fano-like reflectance minimum exhibits only weak $\phi$ dependence, while increasing the groove depth $z$ induces a systematic redshift and modifies the dip depth/linewidth. Spectra of all arrays were measured with the 10$\times$/0.25 objective.}
\end{figure*}

Illumination was configured in a K\"ohler geometry, as shown in Fig.~\ref{SEM-image}(b), with the condenser aperture stop used to control the illumination cone and the field stop used to confine the illuminated region to the metasurface area. Consequently, the measured reflectance spectra represent an average over a circular widefield spot whose diameter decreases with increasing objective magnification (5$\times$ $\rightarrow$ 20$\times$), thereby reducing the contribution from surrounding unpatterned Au at higher magnifications (see Fig.~\ref{fig:appendix_spot_and_mesh}(a)). In addition, the finite numerical aperture (NA) implies an angular spread of incident wave vectors up to $\theta_{\max}\simeq \arcsin(\mathrm{NA})$ (i.e., $\theta_{\max}\approx 6.9\degree$ for NA = 0.12, $14.5\degree$ for NA = 0.25, and $23.6\degree$ for NA = 0.40 in air), which can modestly broaden and reshape narrow resonances compared with idealized plane-wave normal-incidence simulations. Our FEM investigation, however, shows that the resonance wavelength of the $wg$-metasurface is stable, while the spectral linewidth and reflection-minimum value vary modestly with $\theta$.

\section{Results and Discussion}
Before acquiring the reflectance spectra of the $wg$-metasurface, we first measured and subtracted the dark spectrum, i.e., the signal recorded in the absence of illumination. This dark spectrum was used to remove background noise that would otherwise compromise the accuracy of the measurements. Subsequently, a reference spectrum was obtained by reflecting light from an external halogen lamp off a protected silver mirror (coated with a $\mathrm{SiO_2}$ layer to prevent oxidation). This reference spectrum ($I_{\mathrm{ref}}$) was used to normalize the metasurface spectra ($I$) to form the resonator spectrum.

\begin{figure*}[t!]
\begin{minipage}{0.45\linewidth}
\centering
\subfloat[]{\includegraphics[width=1\textwidth]{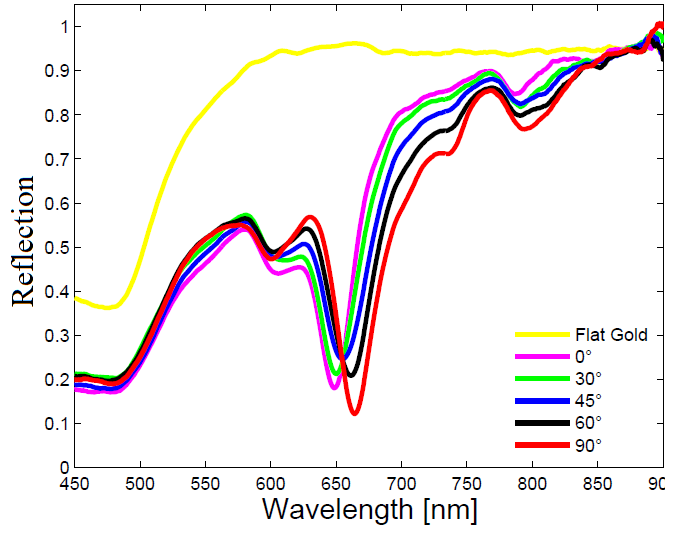}}
\end{minipage}%
\begin{minipage}{0.45\linewidth}
\centering
\subfloat[]{\includegraphics[width=1.03\textwidth]{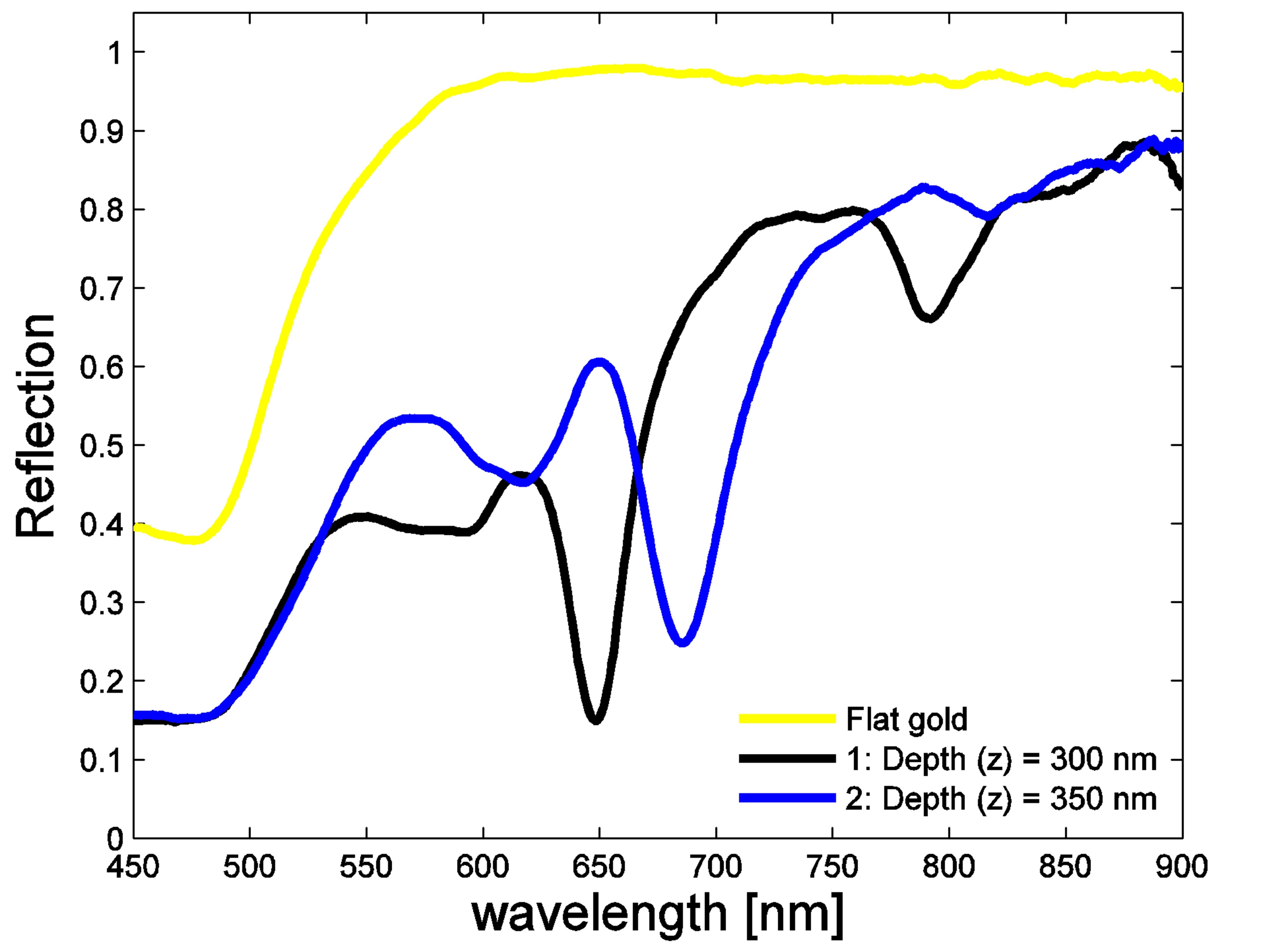}}
\end{minipage}%
\caption{\label{measured_spectra} Optical probing of the two metasurfaces at various azimuthal angles $\phi = 0$--$90\degree$. (a) Measured spectra of the $wg$-metasurface ($D = 500$~nm, $z = 300$~nm, $w = 160$~nm, and $p = 700$~nm). (b) Increasing the groove depth from $z = 300$~nm to $z = 350$~nm induces a spectral redshift of 40~nm. The unstructured flat-gold spectrum is plotted as a background spectrum.}
\end{figure*}

\begin{figure*}[t!]
\begin{minipage}{0.49\linewidth}
\centering
\subfloat[]{\includegraphics[width=1\textwidth]{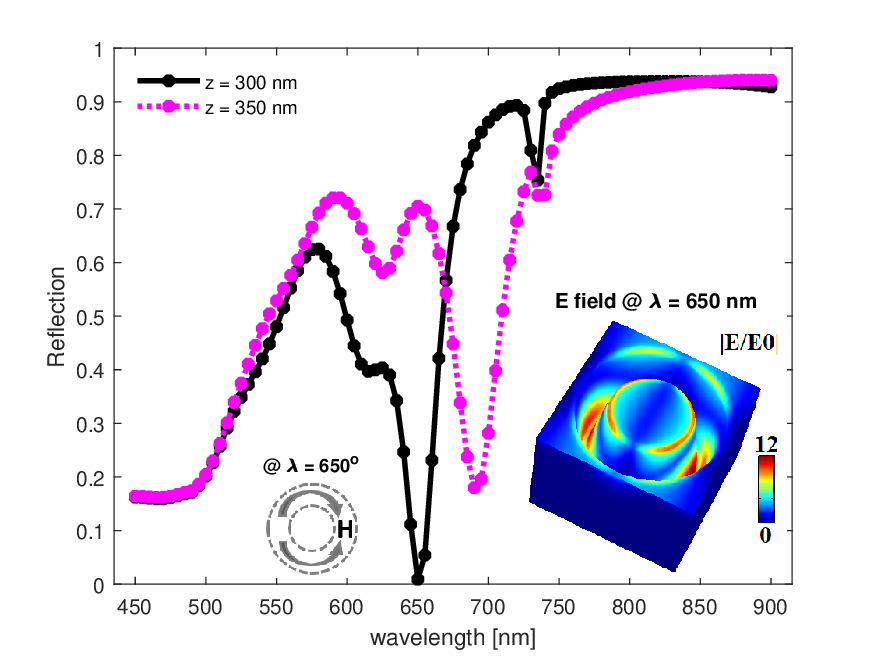}}
\end{minipage}
\begin{minipage}{0.49\linewidth}
\centering
\subfloat[]{\includegraphics[width=1\textwidth]{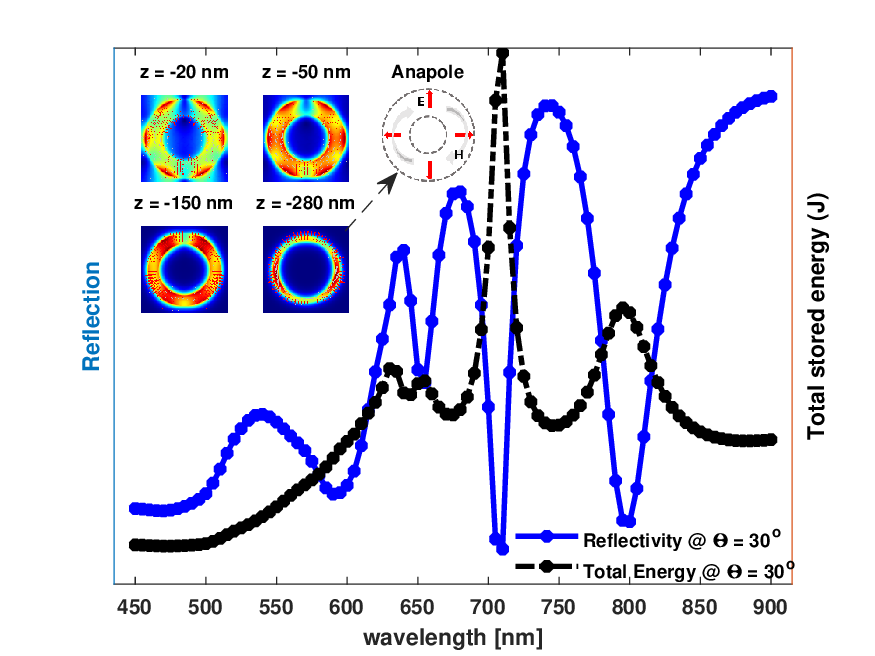}}
\end{minipage}
\caption{\label{analysis1} Optical response of the square-lattice $wg$-metasurface. (a) Calculated spectra of the $wg$-metasurface ($p = 700$~nm, $D = 500$~nm, $z = 300$~nm and 350~nm, $w = 160$~nm, and $t_{Au} = 500$~nm). The near-field distribution and a pictorial representation of the magnetic-field direction $\mathbf{H}$ at $\lambda = 650$~nm inside the groove are shown in the inset. Dipole modes near the bottom ($z = -280$~nm) and octupole-like modes near the top surface ($z = -20$~nm). (b) Anapole-like excitation at $\lambda = 710$~nm when the illumination angle is $30\degree$. Insets show the electric- and magnetic-field distributions at different groove heights measured from the top gold surface.}
\end{figure*}

The silver mirror (purchased from Thorlabs, Inc.) exhibits high reflectance ($>97.5\%$) in the visible range. We also measured the reflectance of unstructured, flat gold, which served as the background spectrum to highlight the influence of $gsp$ excitation on the reflectance of the structured gold (i.e., the resonator array). The differences among the reference spectrum, the background spectrum, and the resonator spectrum are presented in Fig.~\ref{appendix}. Figure~\ref{appendix}(a) is further discussed in Fig.~\ref{measured_spectra}(b) and compared with the FEM calculation in Fig.~\ref{analysis1}. The spectrum of unstructured flat gold was not employed as the reference because of its stronger attenuation (interband transitions) in the 400--600~nm range.

Plasmonic metal templates obtained by conventional deposition techniques, which are usually polycrystalline, introduce additional optical losses originating from grain boundaries and defects. When modified by top-down techniques such as FIB, they can also exhibit increased inhomogeneity~\cite{melngailis1987focused,duan2022recent}. The synthesized Au(111) microplates exhibit high crystalline order, with thicknesses of $\sim 3$--4~$\mu$m and lateral dimensions greater than $10~\mu$m~\cite{radha2012giant}, as shown in Fig.~\ref{SEM-image}(a). Their thickness is much larger than the skin depth of gold at optical frequencies~\cite{maier2007plasmonics}; thus, inscribed V-nanocavities can be treated as carved into an effectively infinitely thick gold film~\cite{maier2007plasmonics,bozhevolnyi2006effective}. The fabricated metasurface therefore achieves spectral selectivity by balancing absorption and reflection at the structured gold surface~\cite{maier2007plasmonics,bozhevolnyi2006effective}.

Single-crystal gold microplates are important for our study because they offer lower optical losses than polycrystalline plasmonic materials~\cite{kaltenecker2020mono, duan2022recent,radha2012giant}. Figure~\ref{measured_spectra} shows the measured reflectance response of the metasurface. Coupling of light to gold plasmons leads to a stable Fano-type resonance with a narrow linewidth: 20~nm wide for $z = 300$~nm and 25~nm for $z = 350$~nm~\cite{cetin2012fano,Feng2009}. The main reflection dip at resonance, $\lambda_{\min} = 650$~nm, reaches a minimum reflectance of $\sim 0.1$--0.2 and is insensitive to changes in the azimuthal viewing angle $\phi$~\cite{zhang2011continuous}. Azimuthal invariance was quantified by tracking the resonance wavelength as a function of sample rotation, yielding a maximum shift of $\max\!\left|\Delta\lambda_{\min}(\phi)\right|<5$~nm (and a change in dip reflectance of $<0.025$) over $\phi=0$--$90\degree$, indicating weak dependence on the azimuthal angle within experimental uncertainty~\cite{zhang2011continuous}. Zhang \textit{et al.}~\cite{zhang2011continuous} attributed this $\phi$ invariance to localized $gsp$ modes of the nanocavities rather than collective diffraction effects of the whole metasurface~\cite{bozhevolnyi2006effective,bozhevolnyi2008scaling,smith2015gap}. The small shift of the main resonance to $\lambda_{\min} = 650\pm 3$~nm (e.g., Fig.~\ref{appendix}(a)) is attributed to the square symmetry of the array~\cite{zhang2011continuous}.

An array with more resonators, such that the incident-light spot captures the entire metasurface area, would improve the results and further maintain $\phi$ invariance, as described in~\cite{zhang2011continuous} and suggested by the high-period $wg$-metasurface in Fig.~\ref{appendix}(d). We also show in Fig.~\ref{measured_spectra}(b) that the metasurface is highly sensitive to the groove depth~\cite{bozhevolnyi2008scaling,vesseur2009modal}. Engineered nanostructures exhibiting these non-natural features have roles in nanotechnology, including color filtering and coatings~\cite{miyata2016full, clausen2014plasmonic, roberts2014subwavelength,zhang2011continuous}. A narrow low-reflectance band can also find applications in anti-counterfeiting optical-variable devices to create high-security reflective patterns, as well as in object-cloaking concepts. Measured spectra of the $wg$-metasurface are compared with FEM (COMSOL Multiphysics) calculated spectra in Fig.~\ref{analysis1}(a).

Figure~\ref{analysis1}(a) compares the calculated spectra at $z = 300$~nm and $z = 350$~nm (using Floquet periodic boundary conditions) with the measured spectra in Fig.~\ref{measured_spectra}(b)~\cite{wong2013high}. The calculated spectra predict the measured spectral profile (Fano-type) and reproduce the major reflection dips at $\lambda_{\min} = 650$~nm and $\lambda_{\min} = 690$~nm~\cite{cetin2012fano,Feng2009}. Overall, the simulated spectra match the measured spectral profile reasonably well. Some divergence occurs because the off-normal incidence components introduced by the objective were not considered in the simulation~\cite{wong2013high}; a normally incident plane wave was used~\cite{wong2013high}. The finite array size may also contribute to the discrepancy. The near-field distribution inside the groove shows a mixture of quadrupolar and strongly symmetric dipolar modes at the top and bottom of the V-cavities~\cite{vesseur2009modal,vesseur2011plasmonic}.

Interestingly, the resonance at $\lambda = 650$~nm corresponds to the excitation of an $m=2$ resonance (azimuthal mode) that has a quadrupolar charge distribution in the plane of the ring resonator, and two antinodes corresponding to a radial mode number $n=2$ as viewed from above~\cite{vesseur2009modal,vesseur2010broadband}. We conclude that coupling between dipolar and quadrupolar modes results in the Fano-type profile seen in Fig.~\ref{measured_spectra}~\cite{cetin2012fano, Feng2009, vesseur2011plasmonic}. The metasurface strongly localizes $gsp$ modes to the groove bottom and edges, with a near-field enhancement ($E/E_0$) of up to 12~\cite{bozhevolnyi2006effective,smith2015gap}, consistent with Eq.~\eqref{wave_guide}. The other resonance at $\lambda = 740$~nm (simulation) and 790~nm (experiment) results from excitation of propagating plasmons at the side edges of the groove walls~\cite{bozhevolnyi2006effective, smith2015gap}.

We next examine the simulated response of the circular V-groove metasurface under oblique incidence to identify regimes of radiatively suppressed scattering accompanied by strong near-field energy confinement~\cite{li2014excitation, miroshnichenko2015nonradiating,savinov2019optical}. In this context, our finite-element calculations indicate an anapole-like excitation near $\lambda \approx 710$~nm for $\theta \approx 30\degree$, manifested by a pronounced reflection minimum together with enhanced electromagnetic energy stored within the cavity volume~\cite{li2014excitation, miroshnichenko2015nonradiating}. We discuss this behavior using the electric- and toroidal-dipole framework commonly employed to describe nonradiating current configurations; however, in the present work, the identification is based on qualitative near- and far-field signatures extracted from the simulations, and a full multipolar decomposition of the radiated power is left for future study~\cite{li2014excitation, miroshnichenko2015nonradiating,savinov2019optical}.

In the electric-dipole approximation, the far-field scattered field may be written as a superposition of contributions associated with the Cartesian electric dipole moment ${P}_{\mathrm{car}}$ and the toroidal dipole moment ${T}_{\mathrm{car}}$, which can destructively interfere in the radiation zone~\cite{evlyukhin2016optical,savinov2019optical}. For a time-harmonic current distribution ${J}(\mathbf{r})$, these moments are defined as~\cite{evlyukhin2016optical}
\begin{equation}
\vec{P}_{\mathrm{car}} = \frac{i}{\omega}\int \vec{J}\,d^3r,
\end{equation}
and
\begin{equation}
\vec{T}_{\mathrm{car}} = \frac{1}{10c}\int\left[(\vec{r}\cdot\vec{J})\vec{r} - 2r^2\vec{J}\right]\,d^3r,
\end{equation}
and an anapole state corresponds to a near-cancellation of the radiated fields associated with these two channels~\cite{miroshnichenko2015nonradiating,savinov2019optical}.

Finite-element simulations at oblique incidence ($\theta \approx 30\degree$) reveal a pronounced reflection minimum near $\lambda \approx 710$~nm, accompanied by strong localization of the electromagnetic energy within the V-groove cavity volume (Fig.~\ref{analysis1}(b))~\cite{li2014excitation,miroshnichenko2015nonradiating}. The simultaneous appearance of a deep far-field dip and enhanced near-field energy storage is consistent with an anapole-like excitation, in the sense of a radiatively suppressed response coexisting with substantial internal field enhancement~\cite{miroshnichenko2015nonradiating,savinov2019optical}. Moreover, insets of Fig.~\ref{analysis1}(b), taken at different heights of the groove profile, show how the electromagnetic near field is distributed around $\lambda = 710$~nm~\cite{li2014excitation}. The fields are characterized by a circulating magnetic field (head-to-tail orientation---often described as poloidal currents) and a radial electric field perpendicular to the cavity circumference. The combination of these two fields forms a key characteristic of the anapole mode---a nonradiating source~\cite{li2014excitation,savinov2019optical}.

In this work, we use an operational definition of an anapole-like regime based on two co-occurring signatures in the numerical model: (i) a pronounced minimum in the far-field reflectance spectrum under oblique incidence, and (ii) a concurrent increase in the electromagnetic energy stored within the cavity volume~\cite{li2014excitation,miroshnichenko2015nonradiating,savinov2019optical}. Specifically, the reflectance dip near $\lambda \approx \SI{710}{\nano\metre}$ at $\theta = 30\degree$ in Fig.~\ref{analysis1}(b) coincides with enhanced field confinement inside the groove, consistent with a radiatively suppressed, energy-storing resonance~\cite{li2014excitation,miroshnichenko2015nonradiating}. We emphasize that definitive identification of a true anapole state requires a multipolar decomposition of the induced currents and radiated power (e.g., demonstrating destructive interference between electric- and toroidal-dipole contributions); such an analysis is beyond the scope of the present study and is left for future work~\cite{evlyukhin2016optical,savinov2019optical}.

The anapole state was first introduced by Zeldovich~\cite{zel1958electromagnetic} and later experimentally demonstrated by~\cite{miroshnichenko2015nonradiating} in high-index silicon nanodisks. Excitation of anapole states is currently a highly active research topic in nanophotonics and plasmonics~\cite{yang2019nonradiating, koshelev2019nonradiating, yao2022plasmonic,savinov2019optical}. It has been explored in contexts including magnetoelectric responses and applications such as Raman enhancement, thermoplasmonics, nanoscale heating, and refractive-index sensing~\cite{zhang2020anapole, baryshnikova2019optical}.

\section{Bulk sensing of analytes}
\begin{figure*}[t!]
\begin{minipage}{0.49\linewidth}
\centering
\subfloat[]{\includegraphics[width=1.00\textwidth]{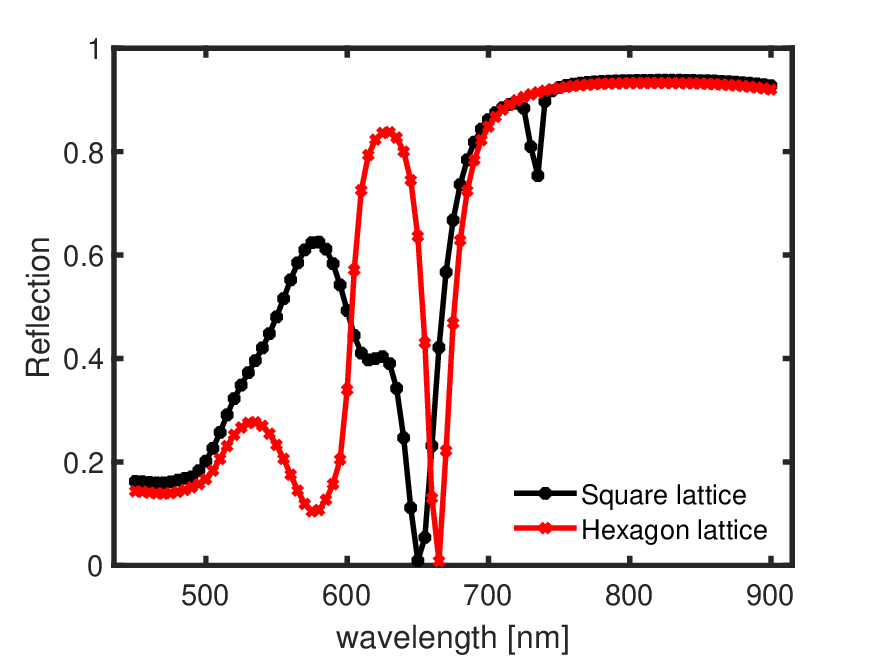}}
\end{minipage}
\begin{minipage}{0.49\linewidth}
\centering
\subfloat[]{\includegraphics[width=1.00\textwidth]{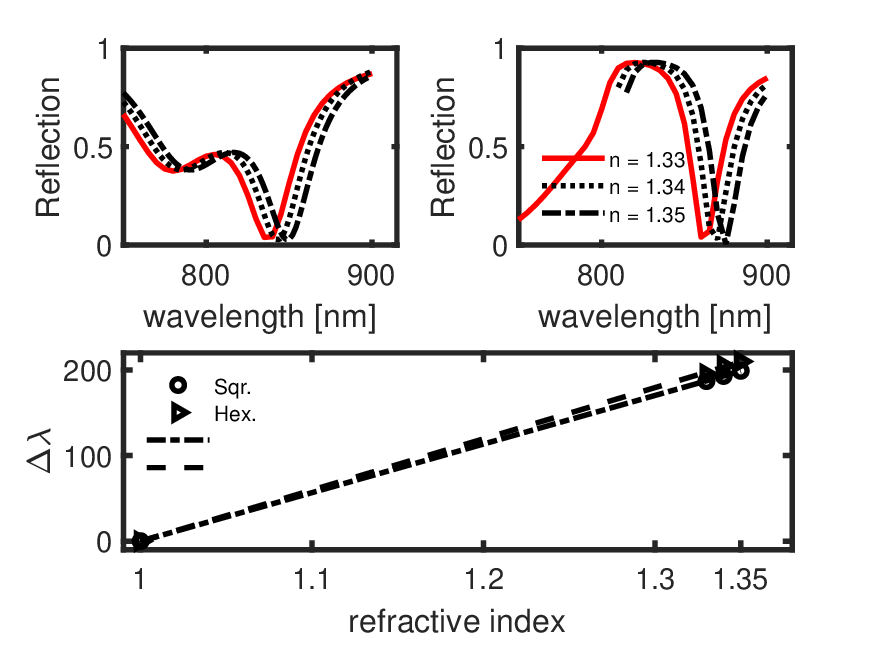}}
\end{minipage}
\caption{\label{hex_spectra} Bulk sensing of the $wg$-metasurface ($D = 500$~nm, $z = 300$~nm). Spectral response of the metasurface for (a) a square-lattice array and (b) a hexagonal-lattice array for four refractive indices $n = 1.00, 1.33, 1.34, 1.35$. (c) Wavelength shift ($\Delta\lambda$) plotted against the refractive index $n$ of the surrounding dielectric medium. $\Delta\lambda = \lambda^{n}_{\min} - \lambda^{0}_{\min}$, where $\lambda^{n}_{\min}$ is the resonance wavelength at $n = 1.33, 1.34, 1.35$ and $\lambda^{0}_{\min}$ is the resonance wavelength at $n = 1.00$. Bulk spectral sensitivities of 567 and 598~nm~RIU$^{-1}$ are obtained for the square- and hexagonal-lattice arrays, respectively.}
\end{figure*}

The $wg$-metasurface also exhibits high sensitivity to small variations in the refractive index of the surrounding medium~\cite{homola2008surface,anker2008biosensing}. Using FEM simulations, we demonstrate the potential of this structure for refractometric sensing by analyzing two different lattice configurations: a square-lattice array and a hexagonal-lattice array of the $wg$-metasurface~\cite{mayer2011localized,piliarik2009surface,vspavckova2016optical}. We evaluate the spectral sensitivity by monitoring the shift in the wavelength position of the principal reflection minimum, $\lambda_{\min}$, for refractive indices $n = 1.00, 1.33, 1.34,$ and $1.35$, as shown in Fig.~\ref{hex_spectra}. The spectral sensitivity of a plasmonic sensor is defined as $S_\lambda = \delta\lambda / \delta n$~\cite{homola2008surface}. In addition, we compute and compare the figure of merit (FOM) for each array. The FOM is defined as $S_\lambda / \Delta\lambda$, where $\Delta\lambda$ is the FWHM of the reflection dip~\cite{vspavckova2016optical,mayer2011localized}. The FWHM characterizes the spectral width of a resonance feature (peak or dip) at half of its maximum intensity~\cite{vspavckova2016optical}.

In Figs.~\ref{hex_spectra}(a) and (b), the linewidths $\Delta\lambda$ for $z = 300$~nm are 20~nm for the square lattice and 18~nm for the hexagonal lattice, respectively. The spectral shift $\delta\lambda$ of the dominant reflection dip---for example, when the refractive index changes from $n = 1.00$ (air) to $n = 1.33$ (water)---is 195~nm for the square array and 187~nm for the hexagonal array. As shown in Fig.~\ref{hex_spectra}(c), the hexagonal lattice configuration attains a higher spectral sensitivity of $598~\mathrm{nm\,RIU^{-1}}$ compared with $567~\mathrm{nm\,RIU^{-1}}$ for the square lattice and consequently exhibits a higher FOM ($\approx 33$), which is approximately 1.2 times larger than that of the square-lattice array ($\approx 28$). Hexagonal lattices generally promote a more closely packed arrangement of resonators than square lattices; however, the overall spectral response of the metasurface is predominantly governed by the optical properties of individual resonators~\cite{zhang2011continuous,vesseur2009modal}.

The relatively small discrepancies in spectral sensitivity and FOM between the two lattice types indicate that inter-resonator coupling remains weak for the selected geometry and periodicity and that the optical response is primarily dictated by near-field confinement within individual V-groove cavities~\cite{vesseur2009modal,vesseur2010broadband}.
These properties provide critical insights for the design of the $wg$-immunosensor. For instance, \textit{Plasmodium falciparum}, the protozoan parasite responsible for malaria, depends on host hemoglobin (Hb) catabolism throughout its intraerythrocytic developmental cycle~\cite{lee2011development,jain2014potential,moody2002rapid}. Because Hb is the main component of the red blood cell (RBC) cytosol, changes in Hb concentration directly cause measurable changes in the RBC refractive index~\cite{park2008refractive,kim2014high}.

Quantitative measurements indicate that the mean Hb concentration decreases from $30.9 \pm 3.1~\mathrm{g\,dL^{-1}}$ in uninfected RBCs to $29.3 \pm 2.4~\mathrm{g\,dL^{-1}}$ at the ring stage, $23.3 \pm 2.7~\mathrm{g\,dL^{-1}}$ at the trophozoite stage, and $18.7 \pm 2.9~\mathrm{g\,dL^{-1}}$ at the schizont stage. These variations correspond to mean refractive indices of 1.399, 1.395, 1.383, and 1.373, respectively~\cite{park2008refractive,kim2014high}. The associated stage-dependent refractive-index changes ($\delta n \approx 0.004$--0.026 relative to uninfected RBCs) are, in principle, detectable by the proposed $wg$-metasurface platform, which exhibits a bulk sensitivity of 598~nm~RIU$^{-1}$~\cite{homola2008surface}.

Provided that the post-functionalization resonance linewidth and the uncertainty associated with determining the dip wavelength remain sufficiently small, the predicted response is expected to produce measurable spectral shifts capable of discriminating infected from uninfected samples and, in principle, enabling refractive-index-based discrimination of infection stage~\cite{homola2008surface,mayer2011localized}. It should be emphasized, however, that red blood cells are structurally heterogeneous and strongly scattering objects; bulk refractive-index sensing of homogeneous solutions is therefore not directly equivalent to whole-blood measurements, and experiments performed with blood lysate may yield outcomes that deviate from the simulation results reported herein~\cite{piliarik2009surface}. The FOM characterizes sensing performance by normalizing the spectral sensitivity to the resonance linewidth; thus, a larger FOM corresponds to a spectrally sharper resonance and improved refractive-index resolution~\cite{vspavckova2016optical}. The calculated FOM (33) is approximately seven times higher than the FOM (4.8) of a comparable localized $wg$-plasmonic sensor ($D = 520$~nm, $z = 375$~nm) reported in~\cite{Vesseur2011PhD}. Consequently, this $wg$-metasurface, implemented as a periodic array of $wg$ resonators, is more suitable as an ultrasensitive sensing platform than individual single $wg$ resonators, and it enables the use of an ultralow detection volume compared with conventional planar SPR sensors~\cite{homola2008surface,anker2008biosensing}.

In contrast to conventional evanescent-field-based surface plasmon resonance (SPR) sensors~\cite{homola2008surface}, the proposed $wg$-metasurface sensor exhibits a Fano-type, spectrally sharp resonance profile~\cite{cetin2012fano}, indicative of a strong, high-contrast optical resonance and, consequently, increased potential quality factor ($Q$ factor) for sensing applications. Such a pronounced resonance is a prerequisite for achieving high sensor sensitivity and a correspondingly improved limit of detection~\cite{mayer2011localized,vspavckova2016optical}.
\begin{figure*}[t!]
\begin{minipage}{0.49\linewidth}
\centering
\subfloat[]{\includegraphics[width=1.0\textwidth]{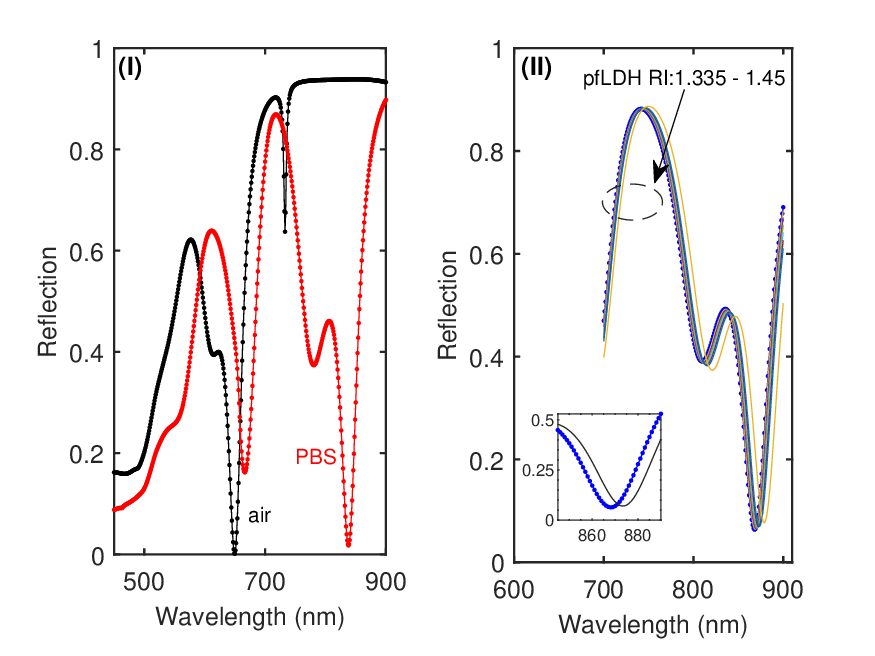}}
\end{minipage}
\begin{minipage}{0.49\linewidth}
\centering
\subfloat[]{\includegraphics[width=1.0\textwidth]{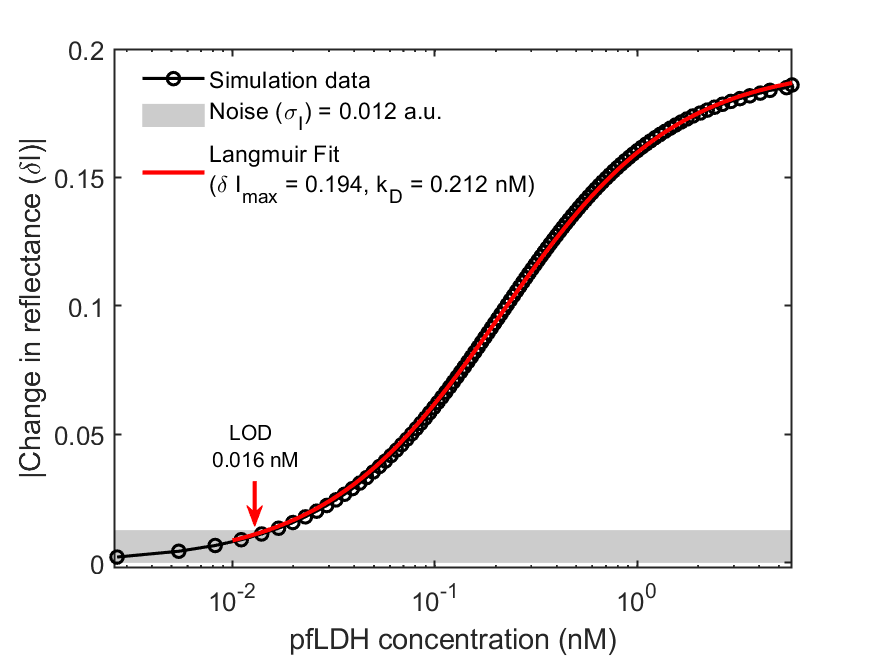}}
\end{minipage}
\caption{\label{model} Functionalized and nonfunctionalized sensor. (a) Reflectance spectra of (I) a nonfunctionalized sensor and (II) a functionalized sensor. The inset in panel (II) is a zoomed-in plot of two spectra: the blue line is the functionalized sensor at zero PfLDH concentration ($n_{\mathrm{eff}} = 1.335$, i.e., buffer solution), and the black line is the spectrum when the bulk PfLDH concentration ($L_0$) is $0.4841~\mathrm{ng\,mL^{-1}}$ (or $n_{\mathrm{eff}}$ of the PfLDH bound layer is 1.405). (b) Langmuir fitting model to simulation data showing the change in reflectance ($\delta I$) versus bulk PfLDH concentration (nM), with a Langmuir fit ($\delta I_\text{max} = 0.194$, $k_D = 0.212$~nM) and signal noise $\sigma_I = 0.012$~a.u.\ taken from \cite{kiyumbi2026experimental}. In an experimental context, the noise floor $\sigma_I$ is expected to vary as a function of optical throughput, detector noise characteristics, spectral resolution, and integration time, thereby influencing the LOD.}
\end{figure*}

To quantify the excitation efficiency of the fundamental gap surface plasmon ($gsp$) mode and its radiative response, we evaluated the quality factor $Q$ and the Purcell factor $F_P$ of the $wg$-metasurface. These two figures are widely used to characterize the optical performance and resonant quality of metasurfaces in sensing and lasing applications and are defined as~\cite{purcell1995spontaneous}
\begin{equation}
Q = \frac{1}{\lambda_{\min}\Delta\lambda}\left\vert\lambda_{\min}^2 - 0.25(\Delta\lambda)^2\right\vert
\label{Qfactor}
\end{equation}
and
\begin{equation}
F_P = \frac{3}{4\pi^2}\frac{Q}{V}\left(\frac{\lambda_{\min}}{n_d}\right)^3
\label{Pfactor}
\end{equation}
where $V$ denotes the optical mode volume, $\Delta\lambda$ is the full width at half the reflectance minimum, and $n_d$ is the refractive index of the cavity medium (air in the present case). Equation~\eqref{Qfactor} is consistent with the standard definition $Q = \nu_0/\Delta\nu$ (with $\nu = c/\lambda$), reformulated in terms of a wavelength-based linewidth. In the narrow-resonance limit, $\Delta\lambda \ll \lambda_{\min}$, this expression reduces to the commonly used approximation $Q \approx \lambda_{\min}/\Delta\lambda$~\cite{joannopoulos2008photonic}. The Purcell-factor expression in Eq.~\eqref{Pfactor} follows the original formulation in Ref.~\cite{purcell1995spontaneous}. Over a broad spectral interval ($\lambda = 550$--1500~nm), the mode volumes of $wg$ resonators embedded in a gold film have been reported to lie well below the diffraction limit of $0.125\lambda^3$, reaching values as low as approximately $0.0075\lambda^3$ in the visible regime (380--700~nm)~\cite{vesseur2010broadband}.

The calculated $Q$ factors at $\lambda_{\min} = 650$~nm (square lattice) and $\lambda_{\min} = 670$~nm (hexagonal lattice) are 32.50 and 37.22, respectively, and are effective spectral quality factors extracted from reflectance, not the intrinsic cavity $Q$. Using Eq.~\eqref{Pfactor}, we estimate Purcell factors $F_P$ up to approximately 350 for our metasurface. This value of $F_P$ falls within the range reported for comparable $wg$ nanostructures of similar diameter in Ref.~\cite{vesseur2010broadband}. Moreover, over a broader photon-energy window from 1.0~eV to 1.8~eV (corresponding to wavelengths of approximately 1200~nm to 688~nm), $Q$ factors up to 50 and Purcell factors up to 2000 have been reported for similar $wg$ metasurfaces incorporating straight, narrow sidewalls~\cite{vesseur2010broadband}. The quality-factor values obtained here are therefore adequate for implementing advanced nanophotonic devices, including ultrasensitive sensors and low-threshold nanolasers~\cite{wei2018channel}.

\section{Application: Surface sensing of PfLDH protein}
We model a surface plasmon resonance (SPR) biosensor based on the $wg$-metasurface architecture for the detection of \textit{Plasmodium falciparum} lactate dehydrogenase (PfLDH), a clinically relevant malaria biomarker~\cite{homola2008surface,anker2008biosensing,lee2011development,jain2014potential}. PfLDH is the predominant lactate dehydrogenase protein produced by all \textit{Plasmodium} species, and its concentration in whole blood is approximately proportional to the level of malaria parasitemia~\cite{lee2011development,jain2014potential,moody2002rapid}. The $wg$ nanoplasmonic biosensor is implemented as a stratified assembly of molecular layers formed on the gold V-shaped sidewalls of the metasurface, following the immobilization strategy described in~\cite{jung1998quantitative}. General parameters used to design the malaria biosensor are given in Table~\ref{parameter}.

In Ref.~\cite{kiyumbi2025effective}, we related the effective refractive index of bound PfLDH, $n_{\mathrm{eff}}$ (or any analyte), to its bulk concentration $L_0$ using Maxwell--Garnett effective medium theory. Bound PfLDH is modeled as a thin layer whose effective refractive index varies with $L_0$. We stress that ligand--receptor binding does not change dipole moments; therefore, the PfLDH antibody and antigen retain their respective permittivities in each layer after binding~\cite{jung1998quantitative}. Phosphate-buffered saline (PBS, $n_{\mathrm{soln}} = 1.335$) is used as the buffer to compute refractive indices of PfLDH solutions at different concentrations~\cite{homola2008surface}. The spectral response of the sensor for various PfLDH concentrations is then calculated using COMSOL.

\begin{table}[th!]
\caption{\label{parameter} General parameters used to design the malaria biosensor from a circular V-cavity metasurface. M.W. is an acronym for molecular weight, and R.I for refractive index.}
\begin{center}
\begin{tabular}{{l}{l}{l}{l}{c}{c}{r}}
\br\hline
Type& Layer&R.I.&Thickness &M.W. & $k_D$ [nM] &Ref.\\
\mr\hline
Organic & DSP &1.45 & $1.00$~nm &  &  &~\cite{schmid2006site}\\
 & Protein A &1.45 & $5.00$~nm &  &  &~\cite{grubor2004novel}\\
Analyte & PfLDH & $f(L_0)$ & $7.95$~nm &35~kDa &0.306 &~\cite{jain2014potential}\\
Receptor& IgG1 & 1.45 & $8.40$~nm & &  &~\cite{tan2008nanoengineering}\\
\br\hline
\label{table1}
\end{tabular}
\end{center}
\end{table}

In Fig.~\ref{model}(a) (left), we show a nonfunctionalized sensor that relies solely on bulk refractive-index changes, which are typically smaller and less detectable unless the analyte concentration is high~\cite{homola2008surface,anker2008biosensing}. Functionalization increases the ability of the sensor to detect low analyte concentrations~\cite{homola2008surface,anker2008biosensing}. For this sensor, the model predicts resolvable shifts for a 0.07 change in refractive index of the bound PfLDH layer (equivalent to a 0.45~nM change in the bulk PfLDH concentration), producing a total spectral shift of 6.5~nm in the resonance position, as shown in the inset of Fig.~\ref{model}(a) (right). This large spectral shift is not possible without functionalization.

In Fig.~\ref{model}(b), where the change in reflectance ($\delta I$) increases with PfLDH concentration, the LOD is estimated as the PfLDH concentration at which the Langmuir fit intersects the noise level~\cite{homola2008surface,mayer2011localized}. The noise level is $\sigma_I = 0.012$~a.u., based on our previous experimental data~\cite{kiyumbi2026experimental}, and is used as a representative normalized-intensity noise for similar acquisition and analysis conditions; if the actual experimental noise differs, the LOD must be scaled accordingly. The Langmuir model equation is expressed in Ref.~\cite{ayawei2017modelling, lavin2018determination}. The Langmuir fit parameters are $\delta I_{\text{max}} = 0.194$ and $k_D = 0.212$~nM. The goodness of fit between the simulation data and the Langmuir equation, i.e., the coefficient of determination $R^2$, is 0.9997. Solving for $L_0$, the estimated LOD is approximately 0.016~nM, equivalent to $0.56~\mathrm{ng\,mL^{-1}}$ of PfLDH. This concentration (or $<0.001\%$ parasite density) falls within the asymptomatic malaria regime~\cite{martin2009unified, moody2002rapid}.

The LOD reported here (0.016~nM of PfLDH, equivalent to $0.56~\mathrm{ng\,mL^{-1}}$) is a model-based estimate derived from simulated calibration curves and an experimental reflectance-noise level expressed in the same normalized units as the response $\delta I$~\cite{lavin2018determination,homola2008surface,mayer2011localized}. In practice, the achievable LOD will be governed by the combined uncertainty in (i) spectral acquisition (wavelength sampling and spectrometer resolution), (ii) resonance extraction (fit/model choice and baseline normalization), (iii) temporal drift and repeatability, and (iv) linewidth broadening and resonance-contrast changes induced by surface functionalization and complex sample matrices~\cite{homola2008surface,mayer2011localized,piliarik2009surface,zeng2014nanomaterials}. These factors enter through the effective noise $\sigma$ and the local slope $S$ of the calibration curve; consequently, improved readout stability and narrower post-functionalization linewidths directly translate into lower detectable concentrations~\cite{homola2008surface,mayer2011localized}. We therefore interpret the present LOD as a feasibility indicator for the metasurface platform, with quantitative performance to be established by dedicated experiments using controlled refractive-index standards and clinically relevant matrices.

\section{Conclusion}
We have demonstrated a reflection-mode whispering-gallery metasurface based on periodic arrays of circular V-groove nanocavities milled into optically thick ($\sim\SI{4}{\mu\metre}$) single-crystal Au(111) microplates. Reflection spectroscopy reveals narrow, depth-tunable Fano-like minima, and unit-cell finite-element modeling captures the principal experimental trends, supporting a physical picture in which gap-surface-plasmon confinement at the groove bottom and edges governs the spectral selectivity. The measured resonances exhibit weak azimuthal dependence, quantified by a sub-\SI{5}{\nano\metre} variation in $\lambda_{\min}$ over $\phi=0$--$90\degree$ for the main dip, indicating robust response within experimental uncertainty.

Beyond normal incidence, the numerical results indicate an anapole-like regime under oblique illumination, evidenced by a pronounced far-field reflectance minimum that coincides with increased electromagnetic energy stored inside the cavity. While definitive identification of an anapole state requires a multipolar decomposition of the induced currents and radiated power, the present near-/far-field signatures motivate further study of radiatively suppressed, energy-storing resonances in metallic V-groove platforms.
Finally, we evaluated refractometric sensing feasibility using the simulated spectral response and find bulk sensitivities up to $\sim\SI{598}{\nano\metre\,RIU^{-1}}$ and figures of merit up to $\sim 33$ for a hexagonal lattice geometry, with only modest differences relative to a square lattice, suggesting weak inter-resonator coupling for the chosen period. The reported detection metrics should be interpreted as feasibility indicators: practical performance will depend on post-functionalization linewidth broadening, spectral extraction uncertainty, temporal drift, and the optical complexity of real sample matrices. Overall, circular V-groove metasurfaces on single-crystal Au offer a compact and reproducible reflection platform that combines spectral tunability with strong near-field localization and sensing-relevant responsivity.

Beyond sensitivity metrics, a key practical advantage of the present platform is its manufacturability and scalability: the circular V-groove unit cell is compatible with wafer-level patterning (e.g., nanoimprint or interference lithography followed by anisotropic etching/metal deposition) and with multi-spot readout in reflection, enabling parallelization and internal referencing on a single chip. Such on-chip referencing is particularly important for suppressing drift from temperature and source fluctuations in field-deployable systems, and it provides a clear route to multiplexed assays where multiple capture chemistries and control regions are measured simultaneously.

\section*{Acknowledgments}
A.\ S.\ Kiyumbi thanks DAAD and the African Laser Centre (ALC) for their funding, M.\ R.\ Goncalves for technical support in the laboratory, and G.\ Neusser for support during the FIB fabrication and SEM imaging of our metasurfaces. The experimental part of this study was done at Ulm University, Germany, under the supervision of Prof.\ O.\ Marti, supported by the Tanzania--Germany Postgraduate Training Program (2015--2017). The computational part, an extension of the experimental part, was done at Stellenbosch University, South Africa, under the supervision of Prof.\ M.\ S.\ Tame, supported by an ALC scholarship at the South African Council for Scientific and Industrial Research (CSIR).

\appendix
\section*{Appendix}
\subsection*{Metasurface unit-cell geometry and numerical discretization}
\begin{figure}[t!]
\begin{minipage}{0.66\linewidth}
\centering
\subfloat[]{\includegraphics[width=1\textwidth]{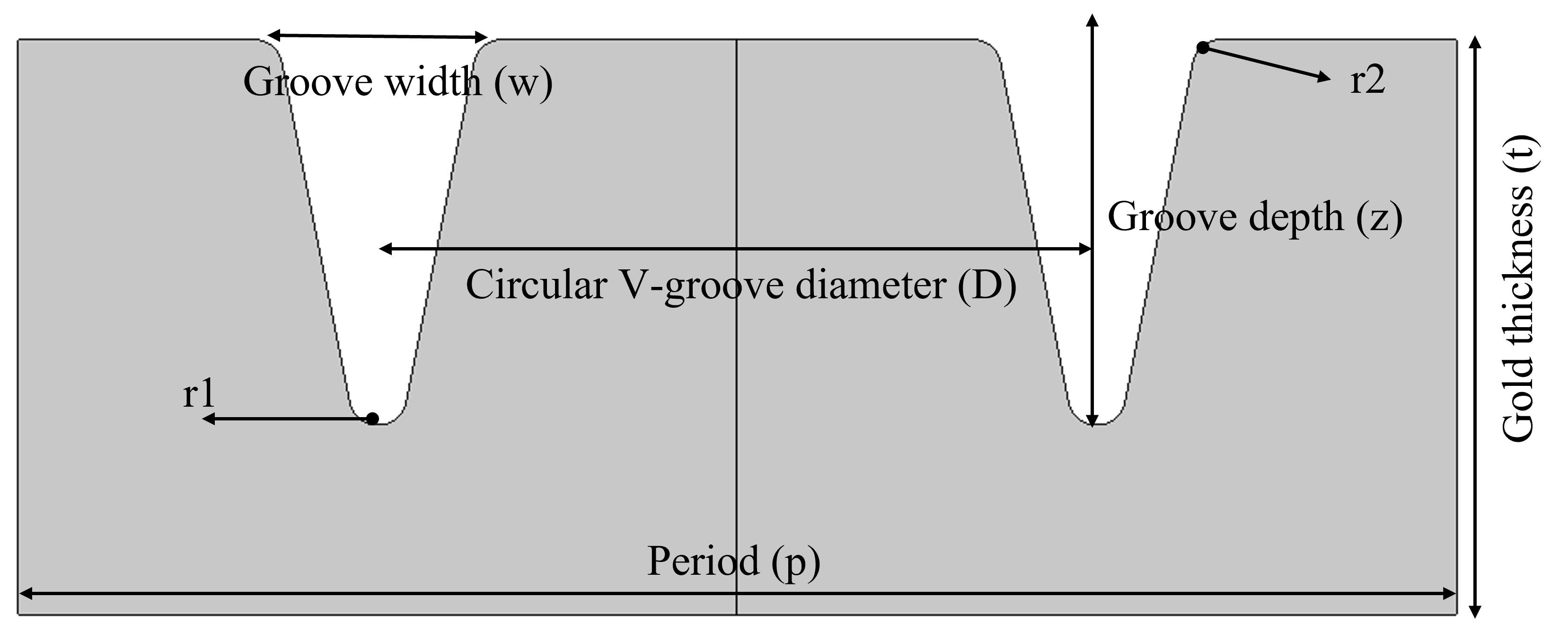}}
\end{minipage}%
\begin{minipage}{0.33\linewidth}
\centering
\subfloat[]{\includegraphics[width=1\textwidth]{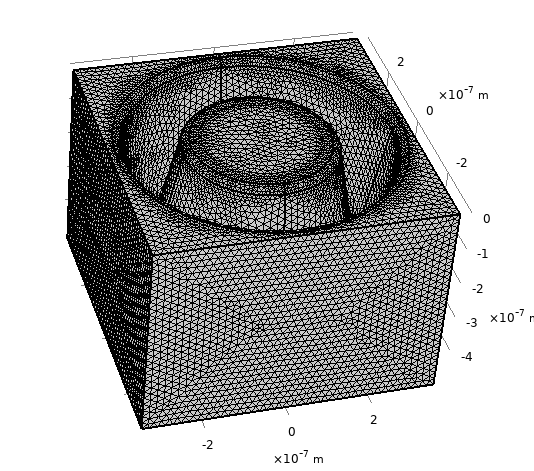}}
\end{minipage}\par\medskip
\caption{\label{design_appendix} The metasurface design and its parameter description. (a) Schematic cross section of the $wg$ resonator. $r_1$ and $r_2$ are minor radii describing the groove curvature at the bottom and top corners, respectively. (b) Meshed unit cell of the $wg$ resonator ($p = 700$~nm, $D = 500$~nm, $z = 300$~nm, $w = 160$~nm, and $t_{Au} = 500$~nm), resembling the FIB-fabricated V-grooves shown in Fig.~\ref{SEM-image}(a)(iv) in the main text.}
\end{figure}

Figure~\ref{design_appendix}(a) summarizes the geometric parameters used to define the circular V-groove resonator and the corresponding finite-element discretization employed in the simulations. The schematic cross section (left) defines the groove width $w$, depth $z$, outer diameter $D$, and array period $p$, together with the inner and outer rounding radii ($r_1$ and $r_2$) that account for the finite tip curvature and edge smoothing produced by focused-ion-beam milling.

These fillet radii are included to avoid unphysical field singularities at perfectly sharp corners and to better represent the fabricated profile. The three-dimensional mesh rendering (right) shows the unit-cell computational domain containing the circular V-groove and surrounding gold volume, discretized with a refined tetrahedral mesh concentrated at the groove sidewalls and bottom, where the electromagnetic fields vary most rapidly. This targeted refinement is essential for accurately resolving gap-surface-plasmon confinement and for obtaining converged reflectance spectra with respect to both geometry and mesh density. The unit-cell geometry was used to study the anapole-like response of the $wg$-metasurface under oblique illumination at $\theta = 30\degree$.

Under normal incidence, the circular cavity and excitation possess mirror symmetries that allow a quarter-cell reduction using PEC/PMC symmetry planes without altering the solution. For oblique illumination, however, the incident field carries a nonzero in-plane wave vector $\mathbf{k}_{\parallel}$, and periodic arrays must satisfy Bloch--Floquet conditions across lattice translations $\mathbf{a}$. Symmetry planes enforce even/odd parity (phase shifts restricted to 0 or $\pi$) rather than the general Bloch phase $e^{i\mathbf{k}_{\parallel}\cdot\mathbf{a}}$, and oblique incidence also breaks the mirror parity of the field components. Consequently, quarter-cell PEC/PMC reductions impose unphysical constraints and can distort resonance positions, linewidths, and coupling strengths; accurate modeling at $\theta\neq 0$ therefore requires the full unit cell, Fig.~\ref{design_appendix}(b), with Floquet-periodic boundaries.

\bibliography{myreferences_revised}

@article{miyata2016full,
  title={Full-color subwavelength printing with gap-plasmonic optical antennas},
  author={Miyata, Masashi and Hatada, Hideaki and Takahara, Junichi},
  journal={Nano letters},
  volume={16},
  number={5},
  pages={3166--3172},
  year={2016},
  DOI={https://doi.org/10.1021/acs.nanolett.6b00500},
  publisher={ACS Publications}
}

@article{kildishev2013planar,
  title={Planar photonics with metasurfaces},
  author={Kildishev, Alexander V and Boltasseva, Alexandra and Shalaev, Vladimir M},
  journal={Science},
  volume={339},
  number={6125},
  pages={1232009},
  year={2013},
  DOI ={10.1126/science.1232009},
  publisher={American Association for the Advancement of Science}
}

@article{bozhevolnyi2006channel,
  title={Channel plasmon subwavelength waveguide components including interferometers and ring resonators},
  author={Bozhevolnyi, Sergey I and Volkov, Valentyn S and Devaux, Eloise and Laluet, Jean-Yves and Ebbesen, Thomas W},
  journal={nature},
  volume={440},
  number={7083},
  pages={508--511},
  year={2006},
  DOI={https://doi.org/10.1038/nature04594},
  publisher={Nature Publishing Group UK London}
}

@article{roberts2014subwavelength,
  title={Subwavelength plasmonic color printing protected for ambient use},
  author={Roberts, Alexander S and Pors, Anders and Albrektsen, Ole and Bozhevolnyi, Sergey I},
  journal={Nano letters},
  volume={14},
  number={2},
  pages={783--787},
  year={2014},
  DOI={https://doi.org/10.1021/nl404129n},
  publisher={ACS Publications}
}

@article{zhang2020anapole,
  title={Anapole mediated giant photothermal nonlinearity in nanostructured silicon},
  author={Zhang, Tianyue and Che, Ying and Chen, Kai and Xu, Jian and Xu, Yi and Wen, Te and Lu, Guowei and Liu, Xiaowei and Wang, Bin and Xu, Xiaoxuan and others},
  journal={Nature communications},
  volume={11},
  number={1},
  pages={3027},
  year={2020},
  DOI={https://doi.org/10.1038/s41467-020-16845-x},
  publisher={Nature Publishing Group UK London}
}

@article{moody2002rapid,
  title={Rapid diagnostic tests for malaria parasites},
  author={Moody, Anthony},
  journal={Clinical microbiology reviews},
  volume={15},
  number={1},
  pages={66--78},
  year={2002},
  DOI={https://doi.org/10.1128/cmr.15.1.66-78.2002
},
  publisher={American Society for Microbiology}
}

@article{vspavckova2016optical,
  title={Optical biosensors based on plasmonic nanostructures: a review},
  author={{\v{S}}pa{\v{c}}kov{\'a}, Barbora and Wrobel, Piotr and Bockov{\'a}, Mark{\'e}ta and Homola, Ji{\v{r}}{\'\i}},
  journal={Proceedings of the IEEE},
  volume={104},
  number={12},
  pages={2380--2408},
  year={2016},
  DOI={10.1109/JPROC.2016.2624340},
  publisher={IEEE}
}

@article{jain2014potential,
  title={Potential biomarkers and their applications for rapid and reliable detection of malaria},
  author={Jain, Priyamvada and Chakma, Babina and Patra, Sanjukta and Goswami, Pranab},
  journal={BioMed research international},
  volume={2014},
  number={1},
  pages={852645},
  year={2014},
  DOI={https://doi.org/10.1155/2014/852645},
  publisher={Wiley Online Library}
}

@article{kiyumbi2025effective,
  title={An effective microscopic model for plasmonic sensing of malaria},
  author={Kiyumbi, AS and Tame, MS},
  journal={Advanced Metamaterials},
  volume={1},
  number={1},
  pages={3},
  year={2025},
  DOI={https://doi.org/10.1007/s44468-025-00003-z},
  publisher={Springer}
}

@article{grubor2004novel,
  title={Novel biosensor chip for simultaneous detection of DNA-carcinogen adducts with low-temperature fluorescence},
  author={Grubor, Nenad M and Shinar, Ruth and Jankowiak, Ryszard and Porter, Marc D and Small, Gerald J},
  journal={Biosensors and Bioelectronics},
  volume={19},
  number={6},
  pages={547--556},
  year={2004},
  DOI={https://doi.org/10.1016/S0956-5663(03)00274-4},
  publisher={Elsevier}
}

@report{WHO2025WMR,
  author      = {{World Health Organization}},
  title       = {World malaria report 2025},
  institution = {World Health Organization},
  year        = {2025},
}

@article{ayawei2017modelling,
  title={Modelling and interpretation of adsorption isotherms},
  author={Ayawei, Nimibofa and Ebelegi, Augustus Newton and Wankasi, Donbebe},
  journal={Journal of chemistry},
  volume={2017},
  number={1},
  pages={3039817},
  year={2017},
  DOI={ https://doi.org/10.1155/2017/3039817},
  publisher={Wiley Online Library},
}

@article{bozhevolnyi2007channelling,
  title={Channelling surface plasmons},
  author={Bozhevolnyi, SI and Volkov, VS and Devaux, E and Laluet, J-Y and Ebbesen, TW},
  journal={Applied Physics A},
  volume={89},
  number={2},
  pages={225--231},
  year={2007},
  DOI={https://doi.org/10.1007/s00339-007-4106-6},
  publisher={Springer}
}

@article{zeng2014nanomaterials,
  title={Nanomaterials enhanced surface plasmon resonance for biological and chemical sensing applications},
  author={Zeng, Shuwen and Baillargeat, Dominique and Ho, Ho-Pui and Yong, Ken-Tye},
  journal={Chemical Society Reviews},
  volume={43},
  number={10},
  pages={3426--3452},
  year={2014},
  DOI={DOI	https://doi.org/10.1039/C3CS60479A},
  publisher={Royal Society of Chemistry}
}

@article{martin2009unified,
  title={Unified parasite lactate dehydrogenase and histidine-rich protein ELISA for quantification of Plasmodium falciparum},
  author={Martin, Samuel K and Rajasekariah, G-Halli and Awinda, George and Waitumbi, John and Kifude, Carolyne},
  journal={The American journal of tropical medicine and hygiene},
  volume={80},
  number={4},
  pages={516--522},
  year={2009},
  DOI={10.4269/ajtmh.2009.80.516},
  publisher={American Society of Tropical Medicine and Hygiene}
}

@article{lavin2018determination,
  title={On the determination of uncertainty and limit of detection in label-free biosensors},
  author={Lav{\'\i}n, {\'A}lvaro and Vicente, Jes{\'u}s de and Holgado, Miguel and Laguna, Mar{\'\i}a F and Casquel, Rafael and Santamar{\'\i}a, Beatriz and Maigler, Mar{\'\i}a Victoria and Hern{\'a}ndez, Ana L and Ram{\'\i}rez, Yolanda},
  journal={Sensors},
  volume={18},
  number={7},
  pages={2038},
  year={2018},
  DOI={https://doi.org/10.3390/s18072038},
  publisher={MDPI}
}

@article{tan2008nanoengineering,
  title={A nanoengineering approach for investigation and regulation of protein immobilization},
  author={Tan, Yih Horng and Liu, Maozi and Nolting, Birte and Go, Joan G and Gervay-Hague, Jacquelyn and Liu, Gang-yu},
  journal={ACS nano},
  volume={2},
  number={11},
  pages={2374--2384},
  year={2008},
  DOI={https://doi.org/10.1021/nn800508f},
  publisher={ACS Publications}
}

@article{schmid2006site,
  title={Site-directed antibody immobilization on gold substrate for surface plasmon resonance sensors},
  author={Schmid, A Hirlekar and Stanca, SE and Thakur, MS and Thampi, K Ravindranathan and Suri, C Raman},
  journal={Sensors and Actuators B: Chemical},
  volume={113},
  number={1},
  pages={297--303},
  year={2006},
  DOI={https://doi.org/10.1016/j.snb.2005.03.018},
  publisher={Elsevier}
}

@article{jung1998quantitative,
  title={Quantitative interpretation of the response of surface plasmon resonance sensors to adsorbed films},
  author={Jung, Linda S and Campbell, Charles T and Chinowsky, Timothy M and Mar, Mimi N and Yee, Sinclair S},
  journal={Langmuir},
  volume={14},
  number={19},
  pages={5636--5648},
  year={1998},
  DOI={https://doi.org/10.1021/la971228b},
  publisher={ACS Publications}
}

@article{park2008refractive,
  title={Refractive index maps and membrane dynamics of human red blood cells parasitized by Plasmodium falciparum},
  author={Park, YongKeun and Diez-Silva, Monica and Popescu, Gabriel and Lykotrafitis, George and Choi, Wonshik and Feld, Michael S and Suresh, Subra},
  journal={Proceedings of the National Academy of Sciences},
  volume={105},
  number={37},
  pages={13730--13735},
  year={2008},
  DOI={https://doi.org/10.1073/pnas.0806100105},
  publisher={National Academy of Sciences}
}

@book{Joannopoulos2008photonic,
  author       = {Joannopoulos, John D. and Johnson, Steven G. and Winn, Joshua N. and Meade, Robert D.},
  title        = {Photonic Crystals: Molding the Flow of Light},
  publisher    = {Princeton University Press},
  year         = {2008},
  edition      = {2}
}

@article{wei2018channel,
  title={Channel plasmon nanowire lasers with V-groove cavities},
  author={Wei, Wei and Yan, Xin and Shen, Bing and Qin, Jian and Zhang, Xia},
  journal={Nanoscale Research Letters},
  volume={13},
  number={1},
  pages={227},
  year={2018},
  DOI={https://doi.org/10.1186/s11671-018-2640-0},
  publisher={Springer}
}

@article{kiyumbi2026experimental,
  title={Experimental plasmonic sensing of malaria using an aluminum metasurface},
  author={Kiyumbi, Amos Sospeter and Tame, Mark},
  journal={Nanoscale Advances},
  pages  ={-},
  year={2026},
  DOI={https://doi.org/10.1039/D5NA01083G},
  publisher={Royal Society of Chemistry}
}

@incollection{purcell1995spontaneous,
  title={Spontaneous emission probabilities at radio frequencies},
  author={Purcell, Edward Mills},
  booktitle={Confined electrons and photons: new physics and applications},
  pages={839--839},
  year={1995},
  DOI={https://doi.org/10.1007/978-1-4615-1963-8_40},
  publisher={Springer}
}

@article{kim2014high,
  title={High-resolution three-dimensional imaging of red blood cells parasitized by Plasmodium falciparum and in situ hemozoin crystals using optical diffraction tomography},
  author={Kim, Kyoohyun and Yoon, HyeOk and Diez-Silva, Monica and Dao, Ming and Dasari, Ramachandra R and Park, YongKeun},
  journal={Journal of biomedical optics},
  volume={19},
  number={1},
  pages={011005--011005},
  year={2014},
  DOI={https://doi.org/10.1117/1.JBO.19.1.011005},
  publisher={Society of Photo-Optical Instrumentation Engineers}
}

@article{lee2011development,
  title={Development and evaluation of a rapid diagnostic test for Plasmodium falciparum, P. vivax, and mixed-species malaria antigens},
  author={Lee, Gyu-Cheol and Jeon, Eun-Sung and Le, Dung Tien and Kim, Tong-Soo and Yoo, Jong-Ha and Kim, Hak Yong and Chong, Chom-Kyu},
  journal={The American journal of tropical medicine and hygiene},
  volume={85},
  number={6},
  pages={989},
  year={2011},
  DOI={10.4269/ajtmh.2011.11-0265},
}

@article{yao2022plasmonic,
  title={Plasmonic anapole metamaterial for refractive index sensing},
  author={Yao, Jin and Ou, Jun-Yu and Savinov, Vassili and Chen, Mu Ku and Kuo, Hsin Yu and Zheludev, Nikolay I and Tsai, Din Ping},
  journal={PhotoniX},
  volume={3},
  number={1},
  pages={23},
  year={2022},
  DOI={https://doi.org/10.1186/s43074-022-00069-x},
  publisher={Springer}
}

@article{koshelev2019nonradiating,
  title={Nonradiating photonics with resonant dielectric nanostructures},
  author={Koshelev, Kirill and Favraud, Gael and Bogdanov, Andrey and Kivshar, Yuri and Fratalocchi, Andrea},
  journal={Nanophotonics},
  volume={8},
  number={5},
  pages={725--745},
  year={2019},
  DOI={https://doi.org/10.1515/nanoph-2019-0024},
  publisher={De Gruyter}
}

@article{yang2019nonradiating,
  title={Nonradiating anapole states in nanophotonics: from fundamentals to applications},
  author={Yang, Yuanqing and Bozhevolnyi, Sergey I},
  journal={Nanotechnology},
  volume={30},
  number={20},
  pages={204001},
  year={2019},
  DOI={10.1088/1361-6528/ab02b0},
  publisher={IOP Publishing}
}

@article{zel1958electromagnetic,
  title={Electromagnetic interaction with parity violation},
  author={Zel’Dovich, Ia B and others},
  journal={Sov. Phys. JETP},
  volume={6},
  number={6},
  pages={1184--1186},
  year={1958},
  url= {http://www.jetp.ras.ru/cgi-bin/dn/e_006_06_1184.pdf},
}

@article{baryshnikova2019optical,
  title={Optical anapoles: concepts and applications},
  author={Baryshnikova, Kseniia V and Smirnova, Daria A and Luk'yanchuk, Boris S and Kivshar, Yuri S},
  journal={Advanced Optical Materials},
  volume={7},
  number={14},
  pages={1801350},
  year={2019},
  DOI={https://doi.org/10.1002/adom.201801350},
  publisher={Wiley Online Library}
}

@article{evlyukhin2016optical,
  title={Optical theorem and multipole scattering of light by arbitrarily shaped nanoparticles},
  author={Evlyukhin, Andrey B and Fischer, Tim and Reinhardt, Carsten and Chichkov, Boris N},
  journal={Physical Review B},
  volume={94},
  number={20},
  pages={205434},
  year={2016},
  DOI={https://doi.org/10.1103/PhysRevB.94.205434},
  publisher={APS}
}

@article{wong2013high,
  title={High throughput and high yield nanofabrication of precisely designed gold nanohole arrays for fluorescence enhanced detection of biomarkers},
  author={Wong, Ten It and Han, Shan and Wu, Lin and Wang, Yi and Deng, Jie and Tan, Christina Yuan Ling and Bai, Ping and Loke, Yee Chong and Da Yang, Xin and Tse, Man Siu and others},
  journal={Lab on a Chip},
  volume={13},
  number={12},
  pages={2405--2413},
  year={2013},
  DOI={DOI	https://doi.org/10.1039/C3LC41396A},
  publisher={Royal Society of Chemistry}
}

@article{clausen2014plasmonic,
  title={Plasmonic metasurfaces for coloration of plastic consumer products},
  author={Clausen, Jeppe S and H{\o}jlund-Nielsen, Emil and Christiansen, Alexander B and Yazdi, Sadegh and Grajower, Meir and Taha, Hesham and Levy, Uriel and Kristensen, Anders and Mortensen, N Asger},
  journal={Nano letters},
  volume={14},
  number={8},
  pages={4499--4504},
  year={2014},
  DOI={https://doi.org/10.1021/nl5014986},
  publisher={ACS Publications}
}

@article{moreno2006channel,
  title={Channel plasmon-polaritons: modal shape, dispersion, and losses},
  author={Moreno, Esteban and Garcia-Vidal, FJ and Rodrigo, Sergio G and Martin-Moreno, L and Bozhevolnyi, Sergey I},
  journal={Optics letters},
  volume={31},
  number={23},
  pages={3447--3449},
  year={2006},
  DOI={https://doi.org/10.1364/OL.31.003447},
  publisher={Optical Society of America}
}

@article{gonccalves2020plasmonic,
  title={Plasmonic resonators: fundamental properties and applications},
  author={Gon{\c{c}}alves, Manuel R and Minassian, Hayk and Melikyan, Armen},
  journal={Journal of physics D: applied physics},
  volume={53},
  number={44},
  pages={443002},
  year={2020},
  DOI={10.1088/1361-6463/ab96e9},
  publisher={IOP Publishing}
}

@article{johnson1972optical,
  title={Optical constants of the noble metals},
  author={Johnson, Peter B and Christy, R-WJPrB},
  journal={Physical review B},
  volume={6},
  number={12},
  pages={4370},
  year={1972},
  DOI={https://doi.org/10.1103/PhysRevB.6.4370},
  publisher={APS}
}

@book{maier2007Plasmonics,
  author    = {Maier, Stefan A},
  title     = {Plasmonics: Fundamentals and Applications},
  publisher = {Springer},
  address   = {New York},
  year      = {2007},
  doi       = {10.1007/0-387-37825-1},
  pages      = {21--52, 141--158, 177--191}
}

@article{Feng2009,
  title={Tunability of subradiant dipolar and quadrupolar plasmon resonances in an asymmetrical nanoparticle dimer},
  author={Hao, Feng and Nordlander, Peter and Sonnefraud, Yannick and Van Dorpe, Pol and Maier, Stefan A},
  journal={ACS nano},
  volume={3},
  number={3},
  pages={643--652},
  year={2009},
  doi = {10.1021/nn900012r},
  publisher={ACS Publications}
}

@article{cetin2012fano,
  title={Fano resonant ring/disk plasmonic nanocavities on conducting substrates for advanced biosensing},
  author={Cetin, Arif E and Altug, Hatice},
  journal={ACS nano},
  volume={6},
  number={11},
  pages={9989--9995},
  year={2012},
  DOI={https://doi.org/10.1021/nn303643w},
  publisher={ACS Publications}
}

@article{melngailis1987focused,
  title={Focused ion beam technology and applications},
  author={Melngailis, John},
  journal={Journal of Vacuum Science \& Technology B: Microelectronics Processing and Phenomena},
  volume={5},
  number={2},
  pages={469--495},
  year={1987},
  DOI={https://doi.org/10.1116/1.583937},
  publisher={American Vacuum Society}
}

@phdthesis{Vesseur2011PhD,
  author       = {Vesseur, Ernst Jan R.},
  title        = {Electron Beam Imaging and Spectroscopy of Plasmonic Nanoantenna Resonances},
  school       = {Utrecht University},
  address      = {Utrecht, The Netherlands},
  year         = {2011},
  month        = jul,
  type         = {Ph.D. thesis},
  url          = {https://www.erbium.nl/wp-content/uploads/2016/08/PhD-thesis-Ernst-Jan-Vesseur-2011.pdf}
}

@article{vesseur2011plasmonic,
  title={Plasmonic whispering gallery cavities as optical nanoantennas},
  author={Vesseur, Ernst Jan R and Polman, Albert},
  journal={Nano letters},
  volume={11},
  number={12},
  pages={5524--5530},
  year={2011},
  DOI={https://doi.org/10.1021/nl203418c},
  publisher={ACS Publications}
}

@article{zhang2011continuous,
  title={Continuous metal plasmonic frequency selective surfaces},
  author={Zhang, Jianfa and Ou, Jun-Yu and Papasimakis, Nikitas and Chen, Yifang and MacDonald, Kevin F and Zheludev, Nikolay I},
  journal={Optics express},
  volume={19},
  number={23},
  pages={23279--23285},
  year={2011},
  DOI={https://doi.org/10.1364/OE.19.023279},
  publisher={Optical Society of America}
}

@article{li2014excitation,
  title={Excitation of plasmon toroidal mode at optical frequencies by angle-resolved reflection},
  author={Li, Jiaqi and Zhang, Yuan and Jin, Renchao and Wang, Qianjin and Chen, Qian and Dong, Zhenggao},
  journal={Optics letters},
  volume={39},
  number={23},
  pages={6683--6686},
  year={2014},
  DOI={https://doi.org/10.1364/OL.39.006683},
  publisher={Optical Society of America}
}

@article{vesseur2010broadband,
  title={Broadband Purcell enhancement in plasmonic ring cavities},
  author={Vesseur, Ernst Jan R and de Abajo, F Javier Garc{\'\i}a and Polman, Albert},
  journal={Physical Review B—Condensed Matter and Materials Physics},
  volume={82},
  number={16},
  pages={165419},
  year={2010},
  DOI={https://doi.org/10.1103/PhysRevB.82.165419},
  publisher={APS}
}

@article{economou1969surface,
  title={Surface plasmons in thin films},
  author={Economou, EN},
  journal={Physical review},
  volume={182},
  number={2},
  pages={539},
  year={1969},
  DOI={https://doi.org/10.1103/PhysRev.182.539},
  publisher={APS}
}

@article{novikov2002channel,
  title={Channel polaritons},
  author={Novikov, IV and Maradudin, AA},
  journal={Physical Review B},
  volume={66},
  number={3},
  pages={035403},
  year={2002},
  DOI={https://doi.org/10.1103/PhysRevB.66.035403},
  publisher={APS}
}

@article{bozhevolnyi2008scaling,
  title={Scaling for gap plasmon based waveguides},
  author={Bozhevolnyi, Sergey I and Jung, Jesper},
  journal={Optics express},
  volume={16},
  number={4},
  pages={2676--2684},
  year={2008},
  DOI={https://doi.org/10.1364/OE.16.002676},
  publisher={Optical Society of America}
}

@article{yerlikaya2022dual,
  title={A dual, systematic approach to malaria diagnostic biomarker discovery},
  author={Yerlikaya, Seda and Owusu, Ewurama DA and Frimpong, Augustina and DeLisle, Robert Kirk and Ding, Xavier C},
  journal={Clinical Infectious Diseases},
  volume={74},
  number={1},
  pages={40--51},
  year={2022},
  DOI={https://doi.org/10.1093/cid/ciab251},
  publisher={Oxford University Press US}
}

@article{platon2026world,
  title={World malaria report 2025: Growing biological threats, shrinking resources},
  author={Platon, Lucien and Lu, Feng and M{\'e}nard, Didier},
  journal={Decoding Infection and Transmission},
  volume={4},
  pages={100076},
  year={2026},
  DOI={https://doi.org/10.1016/j.dcit.2026.100076},
  publisher={Elsevier}
}

@article{mayer2011localized,
  title={Localized surface plasmon resonance sensors},
  author={Mayer, Kathryn M and Hafner, Jason H},
  journal={Chemical reviews},
  volume={111},
  number={6},
  pages={3828--3857},
  year={2011},
  DOI={https://doi.org/10.1021/cr100313v},
  publisher={ACS Publications}
}

@article{piliarik2009surface,
  title={Surface plasmon resonance (SPR) sensors: approaching their limits?},
  author={Piliarik, Marek and Homola, Ji{\v{r}}{\'\i}},
  journal={Optics express},
  volume={17},
  number={19},
  pages={16505--16517},
  year={2009},
  DOI={https://doi.org/10.1364/OE.17.016505},
  publisher={Optical Society of America}
}

@article{yezekyan2022anapole,
  title={Anapole states in gap-surface plasmon resonators},
  author={Yezekyan, Torgom and Zenin, Vladimir A and Beermann, Jonas and Bozhevolnyi, Sergey I},
  journal={Nano Letters},
  volume={22},
  number={15},
  pages={6098--6104},
  year={2022},
  DOI={https://doi.org/10.1021/acs.nanolett.2c01051},
  publisher={ACS Publications}
}

@article{miroshnichenko2015nonradiating,
  title={Nonradiating anapole modes in dielectric nanoparticles},
  author={Miroshnichenko, Andrey E and Evlyukhin, Andrey B and Yu, Ye Feng and Bakker, Reuben M and Chipouline, Arkadi and Kuznetsov, Arseniy I and Luk’yanchuk, Boris and Chichkov, Boris N and Kivshar, Yuri S},
  journal={Nature communications},
  volume={6},
  number={1},
  pages={8069},
  year={2015},
  DOI={https://doi.org/10.1038/ncomms9069},
  publisher={Nature Publishing Group UK London}
}

@article{savinov2019optical,
  title={Optical anapoles},
  author={Savinov, Vassili and Papasimakis, Nikitas and Tsai, DP and Zheludev, NI},
  journal={Communications Physics},
  volume={2},
  number={1},
  pages={69},
  year={2019},
  DOI={https://doi.org/10.1038/s42005-019-0167-z},
  publisher={Nature Publishing Group UK London}
}

@article{kasani2019review,
  title={A review of 2D and 3D plasmonic nanostructure array patterns: fabrication, light management and sensing applications},
  author={Kasani, Sujan and Curtin, Kathrine and Wu, Nianqiang},
  journal={Nanophotonics},
  volume={8},
  number={12},
  pages={2065--2089},
  year={2019},
  DOI={https://doi.org/10.1515/nanoph-2019-0158},
  publisher={De Gruyter}
}

@article{meinzer2014plasmonic,
  title={Plasmonic meta-atoms and metasurfaces},
  author={Meinzer, Nina and Barnes, William L and Hooper, Ian R},
  journal={Nature photonics},
  volume={8},
  number={12},
  pages={889--898},
  year={2014},
  DOI={https://doi.org/10.1038/nphoton.2014.247},
  publisher={Nature Publishing Group UK London}
}

@article{vesseur2009modal,
  title={Modal Decomposition of Surface- Plasmon Whispering Gallery Resonators},
  author={Vesseur, Ernst Jan R and Garc{\'\i}a de Abajo, F Javier and Polman, Albert},
  journal={Nano letters},
  volume={9},
  number={9},
  pages={3147--3150},
  year={2009},
  DOI={https://doi.org/10.1021/nl9012826},
  publisher={ACS Publications}
}

@article{dintinger2009channel,
  title={Channel and wedge plasmon modes of metallic V-grooves with finite metal thickness},
  author={Dintinger, Jos{\'e} and Martin, Olivier JF},
  journal={Optics Express},
  volume={17},
  number={4},
  pages={2364--2374},
  year={2009},
  DOI={https://doi.org/10.1364/OE.17.002364},
  publisher={Optical Society of America}
}

@article{bozhevolnyi2005channel,
  title={Channel plasmon-polariton guiding by subwavelength metal grooves},
  author={Bozhevolnyi, Sergey I and Volkov, Valentyn S and Devaux, Elo{\"\i}se and Ebbesen, Thomas W},
  journal={Physical review letters},
  volume={95},
  number={4},
  pages={046802},
  year={2005},
  DOI={https://doi.org/10.1103/PhysRevLett.95.046802},
  publisher={APS}
}

@article{radha2012giant,
  title={Giant single crystalline Au microplates},
  author={Radha, B and Kulkarni, GU},
  journal={Current science},
  pages={70--77},
  year={2012},
  DOI={https://www.jstor.org/stable/24080388},
  publisher={JSTOR}
}

@article{kaltenecker2020mono,
  title={Mono-crystalline gold platelets: a high-quality platform for surface plasmon polaritons},
  author={Kaltenecker, Korbinian J and Krauss, Enno and Casses, Laura and Geisler, Mathias and Hecht, Bert and Mortensen, N Asger and Jepsen, Peter Uhd and Stenger, Nicolas},
  journal={Nanophotonics},
  volume={9},
  number={2},
  pages={509--522},
  year={2020},
  DOI={https://doi.org/10.1515/nanoph-2019-0362},
  publisher={De Gruyter}
}

@article{duan2022recent,
  title={Recent progress and challenges in plasmonic nanomaterials},
  author={Duan, Huiyu and Wang, Tong and Su, Ziyun and Pang, Huan and Chen, Changyun},
  journal={Nanotechnology Reviews},
  volume={11},
  number={1},
  pages={846--873},
  year={2022},
  DOI={https://doi.org/10.1515/ntrev-2022-0039},
  publisher={De Gruyter}
}

@article{smith2015gap,
  title={Gap and channeled plasmons in tapered grooves: a review},
  author={Smith, Cameron LC and Stenger, Nicolas and Kristensen, Anders and Mortensen, N Asger and Bozhevolnyi, Sergey I},
  journal={Nanoscale},
  volume={7},
  number={21},
  pages={9355--9386},
  year={2015},
  DOI={https://doi.org/10.1039/C5NR01282A},
  publisher={Royal Society of Chemistry}
}

@article{bozhevolnyi2006effective,
  title={Effective-index modeling of channel plasmon polaritons},
  author={Bozhevolnyi, Sergey I},
  journal={Optics express},
  volume={14},
  number={20},
  pages={9467--9476},
  year={2006},
  DOI={https://doi.org/10.1364/OE.14.009467},
  publisher={Optical Society of America}
}

@article{anker2008biosensing,
  title={Biosensing with plasmonic nanosensors},
  author={Anker, Jeffrey N and Hall, W Paige and Lyandres, Olga and Shah, Nilam C and Zhao, Jing and Van Duyne, Richard P},
  journal={Nature materials},
  volume={7},
  number={6},
  pages={442--453},
  year={2008},
  DOI={https://doi.org/10.1038/nmat2162},
  publisher={Nature Publishing Group UK London}
}

@article{homola2008surface,
  title={Surface plasmon resonance sensors for detection of chemical and biological species},
  author={Homola, Ji{\v{r}}{\'\i}},
  journal={Chemical reviews},
  volume={108},
  number={2},
  pages={462--493},
  year={2008},
  DOI={https://doi.org/10.1021/cr068107d},
  publisher={ACS Publications}
}

@article{yu2014flat,
  title={Flat optics with designer metasurfaces},
  author={Yu, Nanfang and Capasso, Federico},
  journal={Nature materials},
  volume={13},
  number={2},
  pages={139--150},
  year={2014},
  DOI={10.1126/science.1232009},
  publisher={Nature Publishing Group UK London}
}

\end{document}